\documentclass[manuscript]{acmart}
\AtBeginDocument{%
  \providecommand\BibTeX{{%
    \normalfont B\kern-0.5em{\scshape i\kern-0.25em b}\kern-0.8em\TeX}}}

\copyrightyear{2023}
\acmYear{2023}
\setcopyright{rightsretained}
\acmConference[CHI '23]{Proceedings of the 2023 CHI Conference on Human Factors in Computing Systems}{April 23--28, 2023}{Hamburg, Germany}
\acmBooktitle{Proceedings of the 2023 CHI Conference on Human Factors in Computing Systems (CHI '23), April 23--28, 2023, Hamburg, Germany}\acmDOI{10.1145/3544548.3580941}
\acmISBN{978-1-4503-9421-5/23/04}
\newcommand{\Modified}[1]{\textcolor{black}{#1}}

\usepackage[toc,page]{appendix}
\usepackage{multirow}
\begin{document}

\title{Understanding Visual Arts Experiences of Blind People}


\author{Franklin Mingzhe Li}
\authornote{Equal contribution}
\affiliation{
  \institution{Carnegie Mellon University}
  \city{Pittsburgh}
  \state{Pennsylvania}
  \country{USA}
}
\email{mingzhe2@cs.cmu.edu}

\author{Lotus Zhang}
\authornotemark[1]
\affiliation{
  \institution{University of Washington}
  \city{Seattle}
  \state{Washington}
  \country{USA}
}
\email{hanziz@uw.edu}

\author{Maryam Bandukda}
\affiliation{ 
  \institution{University College London}
  \city{London}
  \country{United Kingdom}
}
\email{m.bandukda@ucl.ac.uk}

\author{Abigale Stangl}
\affiliation{ 
  \institution{University of Washington}
  \city{Seattle}
  \state{Washington}
  \country{USA}
}
\email{astangl@uw.edu}

\author{Kristen Shinohara}
\affiliation{
  \institution{Rochester Institute of Technology}
  \city{Rochester}
  \state{New York}
  \country{USA}
}
\email{kristen.shinohara@rit.edu}

\author{Leah Findlater}
\affiliation{
  \institution{University of Washington}
  \city{Seattle}
  \state{Washington}
  \country{USA}
}
\email{leahkf@uw.edu}

\author{Patrick Carrington}
\affiliation{%
    \institution{Carnegie Mellon University}
    \city{Pittsburgh}
    \state{Pennsylvania}
    \country{USA}
 }
 \email{pcarrington@cmu.edu}

\renewcommand{\shortauthors}{Li et al.}

%People with visual impairments have a rich and varied relationship with makeup and cosmetics---it offers the potential for self-reinvention and the reshaping of social roles. However, visually impaired individuals may experience barriers to conducting a beauty regime because of the reliance on visual information and color variances in makeup. We present content analyses with 145 YouTube videos to demonstrate the makeup practices of visually impaired individuals, showing their unique practices before, during, and after applying makeup. Based on these makeup practices, we conducted semi-structured interviews with 12 visually impaired people who have experiences in makeup to uncover perceptions on makeup (e.g., blindness is extra pressure in makeup) and challenges in makeup processes (e.g., color differentiation). We further discuss design guidelines and opportunities to support existing practices and help people with visual impairments gain greater independence and joy in their makeup practice (e.g., intelligent system for makeup feedback). Overall, our findings on the unique practices of how visually impaired people do makeup (e.g., differentiate different makeup products) provide knowledge to HCI researchers about the learning process, task completion, and acquiring feedback of blind makeup. The existing challenges we uncovered further brings research opportunities to HCI researchers and product designers on making the whole experience of makeup more accessible to people with visual impairments (e.g., lack of accessible product design).

\begin{abstract}
Visual arts play an important role in cultural life and provide access to social heritage and self-enrichment, but most visual arts are inaccessible to blind people. Researchers have explored different ways to enhance blind people's access to visual arts (e.g., audio descriptions, tactile graphics). However, how blind people adopt these methods remains unknown. We conducted semi-structured interviews with 15 blind visual arts patrons to understand how they engage with visual artwork and the factors that influence their adoption of visual arts access methods. We further examined interview insights in a follow-up survey (N=220). We present: 1) current practices and challenges of accessing visual artwork in-person and online (e.g., Zoom tour), 2) motivation and cognition of perceiving visual arts (e.g., imagination), and 3) implications for designing visual arts access methods. Overall, our findings provide a roadmap for technology-based support for blind people's visual arts experiences.
\end{abstract}





\maketitle

\section{Introduction}
%State of the world
Art allows the expression of important ideas, emotions and beliefs in a multitude of forms, and profoundly influences human society. Experiencing art can bring spiritual satisfaction and self-enrichment to individuals, including people in the blind community~\cite{axel2003art, hayhoe2013expanding}: \textit{``The power of the arts can literally change a person's life by helping him or her develop skills such as leadership, teamwork, communication, self-discipline, and creativity''}, says the president and CEO of American Foundation for the Blind \cite{axel2003art}. 
However, most art is consumed visually (e.g., photography, drawing, painting) and thus poses access barriers for over 2.2 billion population in the world who have vision impairments~\cite{cavazos2021accessible}. Although art museums and galleries increasingly offer accessible tours, these are still limited to a small number of venues and are far from comparably enjoyable to what is offered to sighted visitors. In particular, blind people still face challenges with attending and experiencing visual arts exhibitions independently~\cite{cavazos2021accessible}. 

%Past HCI and accessibility research has explored technology-based approaches to lower these access barriers, including museum navigation support~\cite{ghiani2008supporting,asakawa2019independent,hayhoe2017blind}, audio descriptions~\cite{bernardi2016automatic}, tactile graphics~\cite{mukhiddinov2021systematic}, multimodal feedback~\cite{chase2020pantoguide}, and virtual art tours through smart devices~\cite{ahmetovic2021touch}.

%Appreciation of arts is crucial for blind people to as an art designer when creating arts \cite{axel2003art}.

% The big But
Past HCI and accessibility research has explored technology-based approaches to lower these access barriers, including museum navigation support~\cite{ghiani2008supporting,asakawa2019independent,hayhoe2017blind}, audio descriptions~\cite{bernardi2016automatic}, tactile graphics~\cite{mukhiddinov2021systematic}, multimodal feedback~\cite{chase2020pantoguide}, and virtual art tours through smart devices~\cite{ahmetovic2021touch}. Despite past innovative efforts, we still know little about how people in the blind community adopt existing technology-based supports, and what considerations they make when using these tools to access visual arts. %\Modified{Understanding how blind people adopt different visual art access methods under different contexts, such as locations (e.g., art museums, home), social settings (e.g., with peers, alone), and the goals that the art patron has when encountering visual creative works, would allow HCI researchers to better understand the needs of visual art access for blind people under different contexts and develop customized technological supports to enhance inclusion and inclusive design of visual art experiences for people with disabilities.} For example, visual arts experiences are often considered social activities by blind people~\cite{shinohara2011shadow,bieber2013mind}, and thus it is important to consider how access tools support multiple blind patrons at once. 
\Modified{In particular, while contextual factors, such as social settings, could significantly impact the experience of visual arts appreciation~\cite{shinohara2011shadow,bieber2013mind}, little is known about how blind people adopt visual art access methods under different contexts. We therefore explore blind patrons' visual arts access preferences with considerations of a range of contextual factors---including locations (e.g., art museums, home), social settings (e.g., with peers, alone), personal art appreciation goals and vision conditions---to inform the design of more inclusive and effective technological supports for blind art patrons.}

%and how well a certain tool works with their co-present peers at an art venue is important to consider.
%Furthermore, past work has noted drawbacks of existing visual arts accessibility supports, such as audio descriptions being less effective in conveying spatial information~\cite{cavazos2021accessible,morris2018rich} and tactile graphics being less expressive with complex images ~\cite{kalia2011tactile,brock2015interactivity}.

% Therefore, we did
To explore these problem spaces, we conducted semi-structured interviews with 15 blind visual arts patrons and examined interview insights through a follow-up survey study (\textit{N}=220). We focus on the following research questions throughout the studies:

% (1) What options do blind people currently adopt to access visual arts?  (2) What factors influence blind people's adoption of visual arts access methods?  (3) How do blind people perceive and enjoy visual arts? (4) What technology design opportunities exist in supporting blind people with better visual arts access?

\begin{itemize}
  \item RQ1: How do blind people currently access visual arts?
  \item RQ2: What factors influence blind people's adoption of visual arts access methods, and why?
  \item RQ3: What motivates and brings aesthetic enjoyment to blind people to enjoy visual arts non-visually, and why? 
  \item RQ4: How should technology be developed to support blind people to have better visual arts access?
 \end{itemize}

% The key findings are
From our interviews, we learned about participants' existing adoption of access methods for visual arts appreciation (e.g., visits to art venues, participating in Zoom tours) \textbf{(Section \ref{current practices and challenges})}, their motivation for appreciating visual arts, and how they obtain aesthetic enjoyment from visual arts (e.g., through constructing imaginary artwork based on past visual memories and non-visual cues, through discussing artworks with sighted peers) \textbf{(Section \ref{perception})}. We further extracted eight main design considerations for visual arts access technologies that participants deemed important, including: 1) matching patron comprehension goals with non-visual modes of communication, 2) improving coordination across modalities, 3) balancing between flexibility and consistency, 4) enhancing objective interpretation, 5) establishing shared art vocabulary and grammar, 6) describing visual arts based on individual memory, 7) maintaining synchronous feedback, and 8) destigmatizing accessible art experiences \textbf{(Section \ref{preference})}. From the survey, we quantified blind patrons' motivations and sources of enjoyment for engaging with visual arts, as well as their preferences for different visual arts access methods and over the eight key design considerations uncovered from the interviews, with an exploration of these preferences across different vision conditions \textbf{(Section \ref{survey})}. Finally, we discuss these findings in the context of existing research (e.g., remote art access experiences, art appreciation in social settings and activism, and multimodal approach towards different vision conditions) \textbf{(Section \ref{discussion})}. Overall, we believe our findings and discussion points contribute opportunities to support blind people to better leverage art-related assistive technologies for independent and enjoyable experiences for visual arts appreciation.
%Our research contributes and provides inspiration to existing domain of museum accessibility, tactile design \& 3D fabrication, virtual arts exploration, visual content description, and social stigma \& interaction.
% The contributions of this work are... 

% \begin{itemize}
%   \item RQ1: What are the existing practices around makeup and cosmetics?
%   \item RQ2: What is the importance and perception of doing makeup and cosmetics? And why?
%   \item RQ3: What are the existing challenges around makeup and cosmetics? And how could HCI research contribute to solving challenges with makeup and cosmetics for people with visual impairments?
%   %What are the existing challenges and research opportunities for makeup and cosmetics?
% \end{itemize}

\section{Related Work}
\label{related work}
In this section, we first define visual arts and discuss the significance of visual art appreciation for blind people (Section \ref{visual art for blind people}). \Modified{We then cover current art museum accessibility approaches for blind people (Section \ref{accessible art venue tours}). Finally, we introduce technological innovations of different visual art access methods in-depth (i.e., tactile tools, audio descriptions, interactive multimodal presentations) (Section \ref{existing efforts toward accessible visual art for blind people}).}

\subsection{Visual Arts for Blind People}

\label{visual art for blind people}
% potentially divide this subsection further into two subsubsections - we can decide later
Visual arts is an umbrella term for a broad category of art, including fine arts, contemporary arts, decorative arts, and crafts \cite{VisualAr40:online}. These artworks heavily rely on visual perceptions to encode their meaning, and many assume that blind people do not and cannot enjoy visual arts \cite{axel2003art}. However, blind individuals' interests and desires to visit museums and enjoy visual arts are commonly observed and reported in literature \cite{argyropoulos2015re, buyurgan2009expectations, handa2010investigation, hayhoe2013expanding, axel2003art}. As commented in the resource guidebook for art, creativity, and visual impairments: ``Art Beyond Sight''---``Some have a lust for shape and space and form, just as others are moved deeply by music, or nature'' \cite{axel2003art}, not every blind individual enjoys visual arts, yet everyone ought to be able to access it. Past scholars who promoted art education for blind people provided a set of benefits of engaging in visual arts appreciation, including access to important cultural information, aesthetic enjoyment, and also stimuli to personal development—such as critical thinking skills honed through analyzing and interpreting artworks, cooperative learning obtained from appreciating and creating art in groups, self-awareness, and self-confidence \cite{axel2003art}. In particular, Eisner \cite{axel2003art} proposed that experiencing visual arts could bring positive impact to blind people, by improving Braille reading skills, mobility and map-reading skills, tactile exploration skills, development of texture sensitivity, socialization skills, and improved integration into the local community. 
%All the while, there have been art lovers who later acquired blindness or low vision, and face accessibility challenges in continuing with visual art activities. This group of people includes many of the greatest artists in history: Claude Monet, Mary Cassatt, Pissarro, Degas, Daumier, Renoir, Goya, and so on. In fact, many argue that visual impairments may have enhanced these artists' perspectives, as mental visual activity persists after loss of sight, naturally filling one’s mind with ``imaginations and impressions'' \cite{axel2003art}. 
However, the needs of blind people in visual arts have been long neglected, with ``policy makers and arts managers failing to understand the nature of the need'' and ``educators overprotecting blind students'' from being overwhelmed \cite{axel2003art}. \Modified{Our paper extends efforts to make visual arts appreciation non-visually accessible, by inquiring about blind people's perceptions of the opportunities and limitations of emerging technologies for non-visual engagement with visual arts.}

\subsection{\Modified{Art Museum Accessibility for Blind People}}
\label{accessible art venue tours}

\Modified{Many blind people who are interested in visual arts visit art museums and galleries, which in turn have gradually increased non-visual arts access with ongoing effort from disability activists \cite{asakawa2019independent,urbas2018accessibility,InTheirO37:online}. Being able to physically visit art venues is important to many art patrons, as doing so provides social, cultural, and emotional influences only made possible by being present in the same space as the artwork and being a part of the interaction with other visitors \cite{axel2003art,hayhoe2013expanding}. Recently, there has been increasing legislative effort in the US and Europe toward equal access of art (e.g., Americans with Disability Act \cite{Expandin20:online}, Universal Declaration of Human Rights (Article 27) \cite{assembly1948universal, stamatopoulou2007cultural, Conventi76:online}), and art museums are gradually becoming more accessible to blind people \cite{asakawa2019independent, urbas2018accessibility, InTheirO37:online}. Still, accessibility standards are rarely enforced in art museums and galleries, and merely meeting these standards does not necessarily provide experiences for blind people that meet their expectations \cite{kanter2019let, mesquita2016accessibility, walters2009approaches, urbas2018accessibility}.} 

\Modified{To visit an art museum or gallery, one common approach for blind visitors is through specialized, guided tours, often with a docent or a sighted companion who provides navigation support and describes the artwork \cite{asakawa2019independent,jansson2003new,ForVisit63:online}. However, these tours often require the blind patrons to show up on a specific day or need reservation ahead of time \cite{HowDoPeo18:online,Programs66:online,cavazos2021accessible}. Beyond guided tours, pre-recorded audio descriptions of art pieces have become common in museums and galleries \cite{ginley2013museums}, albeit designed primarily for sighted visitors \cite{asakawa2019independent}. To further support blind patrons with more direct experiences, some museums provide tactile options for blind visitors, such as tactile replicas, graphs, or braille-based brochures and tags \cite{Accessib19:online,AboutUsA46:online,anagnostakis2016accessible,urbas2018accessibility}. In reality, these art programs that provide touchable artwork and tailored descriptions are rare among art museums \cite{cavazos2021accessible}. To support blind people navigating in the museum, prior research has explored location tracking, audio notifications, and voice descriptions to inform blind visitors about obstacles, wayfinding, and descriptions of visual art pieces \cite{asakawa2019independent,rector2017eyes,jain2014pilot,meliones2018blind}. In this paper, we explore blind people's overall experiences with visual arts under different contexts and their preferences for appreciating visual arts.}

%Overall, Asakawa et al. \cite{asakawa2019independent} identified two key barriers to blind visitors having a satisfactory museum experience: the lack of independence and accessibility, and a lack of high-quality descriptions of visual artwork \cite{asakawa2019independent,asakawa2018present,ahmetovic2021touch}. 

%Past work that aimed to use technological innovations to allow independent navigation focused mostly on museums. So far, these solutions have used location tracking, audio notifications and voice descriptions to inform blind visitors about obstacles, wayfinding, and descriptions of visual art pieces \cite{asakawa2019independent,rector2017eyes,jain2014pilot,meliones2018blind}. Challenges with these technologies remain with the presence of other museum visitors blocking the path or gathering around \cite{asakawa2019independent}.

\subsection{\Modified{Technological Support of Visual Art Access for Blind People}}
\label{existing efforts toward accessible visual art for blind people}
%Existing research has investigated technologies to help blind people with different daily activities, such as object identification \cite{bigham2010vizwiz,parlouar2009assistive}, navigation \cite{giudice2008blind,ran2004drishti,williams2013pray,azenkot2011enhancing}, and interaction with different interfaces \cite{guo2016vizlens,kane2011usable,mendes2020collaborative}.
Below, \Modified{we expand on new technological innovations to support non-visual appreciation of visual arts, structured into three categories: tactile tools (Section \ref{tactile tools}), audio descriptions (Section \ref{audio descriptions}), and interactive multimodal presentations (Section \ref{Interactive multimodal presentation})}.

 % Besides independence and inaccessibility of artwork, Ahmetovic et al. additionally identified challenges around obtaining support from art venue staff and accessing information about the venue itself [2]. 

\subsubsection{Tactile Tools}
\label{tactile tools}
Touching is one of the main methods for blind people to access visual arts, such as through tactile graphics—graphics made using raised lines and textures to convey drawings and images \cite{cantoni2018art,holloway2019making,cavazos2021accessible}. Without prior training, blind people can capture the sizes, shapes, and location of essential details through touching outlines of objects, which is often faster than decoding descriptions of such details \cite{kennedy1974psychology,axel2003art}. An experienced tactile graphic consumer can identify patterns of shadow and light in artworks \cite{axel2003art}, and the sense of touch can produce accurate and complete characteristics of objects with training \cite{luo2017robotic}. However, tactile graphics can be less effective in expressing visual information of complex images \cite{baker2014tactile}. Therefore, tactile graphics are suitable for simple graphs and diagrams with objects’ spatial relationship being important \cite{cavazos2021accessible,Guidelin92:online}. Still, there have been efforts toward using tactile graphics for complex visual information, such as color \cite{cho2021tactile}, comics \cite{dittmar2014comics,fraser2021tactile}, and also visual artwork \cite{hinton1991use}. However, producing tactile graphics that effectively present complex visual information requires time and skill from professionals, and thus is mostly unavailable, or low-quality \cite{mesquita2016accessibility,walters2009approaches,handa2010investigation}. Recently, there have been attempts to auto-generate tactile graphics or 3D reproductions through AI algorithms \cite{mukhiddinov2021systematic,neumuller20143d,cavazos2018interactive,Touching59:online,Accessib19:online}. Although the quality of such products is still not always ensured, it has the potential to increase tactile access to art in scale. Other challenges of tactile graphics include inadequate educational resources for blind people to learn to consume tactile arts properly \cite{stangl2019defining}, the difficulty of maintaining tactile arts \cite{mesquita2016accessibility}, stigma and restrictions related to touch \cite{candlin2004don,stangl2019defining,kleege2017more}, and copyright infringement related to reproducing visual media in tactile form \cite{holloway2019making}.

%Artistic experiences and expressions through touch: “All the time I pity those who look at things with their hands in their pockets and do not take the trouble to explore the delights of touch or understand how it ministers to their growth, strength and mental balance”; 
% Please do touch the art [14]

\subsubsection{Audio Descriptions}
\label{audio descriptions}
Audio description is another common way for blind people to learn more about visual art pieces, often pre-recorded on either a personal device or a device prepared by the venue \cite{ahmetovic2021touch,axel2003art,mesquita2016accessibility,de2007intersensorial,haworth2012using}.  Audio descriptions are now available in many museums, but are often designed for sighted visitors, and thus still present accessibility barriers for blind visitors \cite{ahmetovic2021touch}. Audio descriptions can usually provide accurate spatial understanding and details of an artwork \cite{neumuller20143d,cavazos2021accessible}. Comparatively, ``verbal imaging''---art descriptions tailored and read by curators of the art venues for blind people---are more compelling but very rarely found \cite{axel2003art}. One way of improving understanding through audio descriptions is to provide multiple levels of descriptions, ranging from general information, to fine-grained details \cite{ahmetovic2021touch,morris2018rich}. The other way of improving understanding through audio descriptions consults general image description generation techniques \cite{ahmetovic2021touch}. To better improve the quality of auto-generated artwork description, past work has also incorporated eye gaze and discussion data from sighted viewers to create rich descriptions for image areas that people tend to pay attention to \cite{sibert2000evaluation,reingold2003gaze}. Huh et al. \cite{huh2022cocomix} further proposed using comments online to automatically generate image descriptions of webtoons (i.e., digital comics) and allow blind readers to request information in a question-and-answer format. Moreover, there are attempts to use sound design to better present comics instead of only verbal descriptions \cite{BlindAcc0:online}. \Modified{Given these technological explorations on audio descriptions, it remains unknown how blind people adopt different audio description methods, and what their associated perceptions and challenges are when using audio descriptions for visual art access.}

%Provide certain evaluation metrics: easy to access, understandable, cognitively demanding, useful, interesting, captivating, exhaustive [2]
% Mobile devices - touch screen exploration + screen reader, visuo-spatial exploration; Employ crowdsourced figurative art description [2]

\subsubsection{Interactive Multimodal Presentations}
\label{Interactive multimodal presentation}
Because audio descriptions and tactile tools are limited in delivering visual information, there have been increasing proposals to provide multimodal presentations of visual arts for blind people, mostly combining audio descriptions and tactile tools \cite{ahmetovic2021touch,iranzo2019exploring,cavazos2018interactive}. Such a multimodal approach also benefits art interpretation, as the non-visual medium used for presenting visual arts could often influence the message and intention expressed \cite{cavazos2021accessible}.
Multimodal approaches to experiencing visual arts also can influence the message and intent expressed in non-visually describing artworks~\cite{cavazos2021accessible}. A range of system innovations further added interactivity to these multimodal presentations to help blind people understand visual arts through exploration \cite{cavazos2021accessible,cho2021tactile,anagnostakis2016accessible,leporini2020design,vaz2018designing,reichinger2016gesture,hinton1991use}. Such interactivity is argued to be important, as ``The goal of my art is to cause a reaction when someone sees it, they (viewers) should think, they should react. That’s what experiencing art is'' \cite{cavazos2021accessible}. For example, recent innovations allow users to perform specific touch gestures on the artwork surface, lift up an item, or press buttons to get localized descriptions or related audio elements as they go \cite{cavazos2021accessible,anagnostakis2016accessible,vaz2018designing,leporini2020design,reichinger2016gesture,hinton1991use,cho2021study}. Last, a range of virtual, internet-based technologies allow remote visual artwork appreciation for blind people. For example, Howell and Porter \cite{howell2003re} proposed to provide home-printed tactile presentations together with online audio guides of artworks to blind people. 

While prior research has explored various technological approaches to access visual arts, it still remains unknown how these methods are adopted by people in the blind community under different contexts, including their preferences among different access technologies and design implications towards more accessible and enjoyable art experiences. Building upon existing visual arts accessing methods, this paper bridges these gaps.
 %Interactive interface for color recognition, applying patterns, sounds, temperature, and/or scents [22]
%Thermoforming with tape recording & tactile graphics of visual artwork [4k]

% \section{Method}

% To understand the research questions, we first conduct semi-structured interviews with people who are legally or completely blind. Based on the findings from the semi-structured interviews, we then conducted a follow up survey study to further identify design preference of our findings.

\section{Semi-structured Interview}
\label{Semi-structured Interview}
We first conducted semi-structured interviews with people who are legally or totally blind to learn how they currently engage with visual arts, including what methods they adopt to access visual arts, what challenges are encountered in the process, and how they cognitively perceive visual arts. At the end of the interview, we explored factors that influence their preferences for existing technology-based access methods.

\subsection{Participants}
We recruited 15 blind participants (P1 - P15) (Table \ref{table:participants}) through different online platforms (e.g., Twitter), email lists  (e.g., National Federation of the Blind Tactile Art and Tactile Graphics Specialist Group), and newsletters (e.g., Blind Posse Newsletter). To participate in our study, participants needed to be 18 years or older, be legally or totally blind, have experience with going to a specific venue for visual arts appreciation (e.g., museum, art gallery), and be able to communicate in English. Among the 15 participants we recruited, seven of them were female, and eight were male (Table \ref{table:participants}). They had an average age of 36.1 years old (SD = 20.9). Nine of them are legally blind, and six are totally blind. Four of them are congenitally blind, and others acquired blindness at different ages. The interview took around 75 to 90 minutes per participant. Participants were compensated with a \$20 Amazon gift card. The recruitment and study procedure was approved by the Institutional Review Board (IRB).

\def\arraystretch{1.15}
\begin{table*}[t]
\centering
\resizebox{1\textwidth}{!}{%
\begin{tabular}{
p{1.6cm}|
p{1.0cm}
p{0.7cm}
p{2cm}
p{3.1cm}
p{6.5cm}
}
    \toprule
    \textbf{Participant} & 
    \textbf{Gender ~~~~} &
    \textbf{Age ~~~~~~~~} & 
    \textbf{Blindness} & 
    \textbf{Occupation} &
    \textbf{Details}\\
    \midrule
P1 & Female & 27 & Legally blind & Tactile painting creator & Congenitally blind, only has light perception.
 \\
P2 & Female & 29 & Totally blind & Mental healthcare & Acquired blindness around six and a half.
%Gross and fine motor difficulties 
\\
P3 & Male & 25 & Legally blind & Student & Acquired around 20.
 \\
P4 & Female & 27 & Legally blind & Online worker & Acquired since six years old. \\
P5 & Male & 21 & Legally blind & Unemployed & Acquired in high school.\\
P6 & Female & 73 & Totally blind & Writer and editor & Congenitally blind.\\
P7 & Male & 74 & Totally blind & Retired physician & Normal sight at 20, legally blind around 30, lost all vision around 50.\\
P8 & Female & 20 & Legally blind & Student & Acquired since six years old.
 \\
P9 & Male & 80 & Totally blind & Rehabilitation counselor & Congenitally blind, totally blind since 1978. \\
P10 & Female & 28 & Totally blind & Unemployed & Acquired at 13 or 14 years old. \\
P11 & Male & 22 & Legally blind & Student & Acquired at 10 years old.
 \\
P12 & Male & 23 & Legally blind & Therapist & Acquired at 17 years old. \\
P13 & Male & 25 & Legally blind & Consultant & Acquired six years ago, personal accident. \\
P14 & Male & 30 & Legally blind & Computer engineer & Acquired at 20 years old. \\
P15 & Female & 38 & Totally blind & Massage Therapist & Congenitally blind, totally blind at 29 years old \\

    \bottomrule
\end{tabular}%
}
\caption{Participants' demographic information.}
\Description{This is the table of our participants' demographic information. There are six columns and 15 rows (without headers). The headers are participant, gender, age, blindness, occupation, and details.}
\label{table:participants}
\end{table*}

%They had an average age of 31.6 (SD = 8.0). Four participants stated that they had spinal cord injuries, four had cerebral palsy, one had stroke, one had primary lateral sclerosis, one had arthrogryposis multiplex congenita, and one had muscular dystrophy. The study took around 75 to 90 minutes per participant. Participants were compensated with a \$20 Amazon gift card. The recruitment and study procedure was approved by the Institutional Review Board (IRB).

\subsection{Study Procedure}
\subsubsection{Demographic Background [5 Minutes]:}
In our semi-structured interviews, we first asked about the demographic background of our participants, which included age, gender, profession, and descriptions of vision level. 

\subsubsection{Current Practices and Challenges of Visual Arts Appreciation [20 Minutes]:}
We then asked our participants about their experiences with visual arts appreciation, inquiring into where they usually visit to enjoy visual arts, who they enjoy visual arts with, what methods they usually leverage to enjoy visual arts and the challenges they experience when appreciating visual arts. Beyond the overall experiences, we also asked about the practices they used to engage with visual arts after they initially lost their vision (if applicable), as compared to their latest experiences to identify any challenges they experienced at different stages of their art appreciation journeys.

\subsubsection{Perceptions of Visual Arts Appreciation [15 Minutes]:}
We further asked our participants about the goals the art patron has when encountering visual creative works and the meaning of visual arts to them in life. According to the description framework of visual content \cite{stangl2021going}, we also asked about participants' awareness and perceptions of details for different elements of visual arts, such as subject (e.g., people, environment, activity), form (e.g., shape, line, color), and content (e.g., history, emotion). 

\subsubsection{Preferences on Alternative Arts Accesses for Visual Arts [40 Minutes]:}
Afterward, we asked participants to envision their preferred ways to consume visual arts, and then introduced a range of existing methods for presenting visual arts non-visually (i.e., tactile graphics, audio descriptions, a combination of tactile and audio presentation, remote visual arts experiences through smart devices, and companion of sighted peer \& tour) based on prior research and accessible blind art guidelines (e.g., \cite{bernardi2016automatic,mukhiddinov2021systematic,chase2020pantoguide,ahmetovic2021touch}). For each method, we asked participants questions such as whether they had experiences with a certain method, how well the method provides information about the art piece, how intuitive or not intuitive the method is, how well the method provides aesthetic enjoyment, and any challenges associated with the method, and how can the method be improved.

\subsection{Data Analysis}
The semi-structured interviews were conducted through Zoom \cite{VideoCon42:online}, and all interviews were audio-recorded and transcribed. After the interviews, two researchers independently performed open coding \cite{charmaz2006constructing} on the transcripts. They then met to discuss their codes and resolve any conflicts (e.g., missing codes, disagreement on codes). After the two researchers reached a consensus and consolidated the list of codes, they performed affinity diagramming \cite{hartson2012ux} using a Miro board \cite{AnOnline70:online} to cluster the codes and identify emergent themes. 

\section{Findings: Current Practices and Challenges of Visual Arts Appreciation for Blind People}
\label{current practices and challenges}
In this section, we present current practices and challenges of blind visual arts patrons, including their \textit{in-person} and \textit{remote} art appreciation experiences.
%In this section, we will first present the practices and challenges of in-person experiences for art appreciation by blind people, such as the practices they use when going to the same art gallery to recall a past memory, as well as their experiences using different visual art access methods intentionally designed for blind people. We then present the practices and challenges of remote experiences for art appreciation (e.g., Zoom tour).

\subsection{In-Person Experiences}
% in-person experience is the most common way to engage with visual arts. A summary of what they use to engage with, what types of visual arts. 
The most common way for our participants to engage with visual arts is to visit art galleries and museums in person (N=$14$). Participants commonly expressed interest in paintings, drawings, and photography exhibited in galleries and museums, and they primarily rely on four non-visual approaches to access these artworks: descriptions from sighted peers or docents (N=$14$), pre-recorded audio descriptions (N=$9$), tactile graphics (N=$6$), and interactive smart device support (N=$2$). 

% Among all participants, we found 14 out of 15 participants usually go to art galleries to enjoy visual arts (e.g., painting, drawing, photography). 
%In terms of methods to enjoy visual arts in galleries or museums, we learned that our participants normally appreciate visual arts through companions such as sighted peers or docents (N=$14$), audio descriptions (N=$9$), touch on tactile graphics (N=$6$), and individually with smart devices (N=$2$). 

% preferences & challenges with specific non-visual art presentations
Among these approaches, participants strongly preferred \textbf{descriptions from sighted peers or docents}, as they could ask questions and get answers immediately from a real person, as P7 commented:
%\textbf{Companions with sighted people:} Regarding appreciating visual arts through companions with sighted peers and docents, we found that our participants prefer a real person to explain arts to them because they could ask questions and get the answers immediately. P7 commented on the reasons:
\begin{quote}
    ``...Having a person with me allows me to ask whatever questions about different art pieces that I have. I could then receive answers immediately...''
\end{quote}

However, P10 and P14 commented on the learning curve for training sighted people to properly explain visual arts to a blind person, as P10 explained:

\begin{quote}
    ``...My cousin initially took me to the art gallery to cheer me up. I appreciate that, but the initial experience was terrible, I barely got anything from her explanation...I remember she just said that this painting has two trees and a farmer between the trees. It takes her two years to provide the details that I would like to know from paintings...such as the correct color shades, visual references of different objects, and interactions between objects...''
\end{quote}

The second most adopted approach is \textbf{pre-recorded audio descriptions} provided by art galleries or museums. One challenge our participants experienced with existing audio description technologies is the additional effort involved, such as the need to scan QR codes and stand exactly in front of the painting. Existing audio descriptions are also mostly designed for sighted patrons, and thus lack detailed descriptions needed for blind people to understand the visual artwork. P7 commented:

\begin{quote}
    ``...Many audio devices I got from art galleries are designed for sighted people. For example, the descriptions do not typically include any visual references, they mentioned several attributes, like houses and trees, but without including where they are, what colors they were painted...''
\end{quote}

For \textbf{tactile graphics}, our participants experienced two main challenges. First, tactile graphics are not available in many art galleries and museums. P6 talked about her limited experiences with tactile graphics in her 70 years of art experiences:

\begin{quote}
    ``...I have over 70 years experiences of art, I can say that I only had opportunities to try tactile art at the Chicago Art Museum (the Art Institute of Chicago). They are just not common...''
\end{quote}

Second, some participants find it particularly time-consuming to access visual arts through tactile graphics, as P12 explained:

\begin{quote}
    ``...It usually takes 15 to 20 minutes to feel a tactile painting, sometimes, it is even longer depending on the details of the painting...''
\end{quote}

Two participants also mentioned the use of \textbf{smart device supports} (e.g., visual interpretation technologies such as Be My Eyes and Aira) to identify what is in an artwork (e.g., subject matters, activities, shapes, colors). While these tools are versed at explaining visual concepts to blind people, they often lack expertise in visual arts. P15 explained:

\begin{quote}
    ``...I found people of BeMyEyes generally have difficulties expressing artwork properly, such as color shades...''
\end{quote} 

Existing visual interpretation tools are also not specifically designed to provide descriptions for artworks, and thus often miss important contextual information in them, as P7 commented: 

\begin{quote}
    ``...I tried to use some object recognition or image description apps on my phone, it mostly tells me things like there are two people, one sword, and one horse. There is almost no contextual information of the painting, such as who is holding the sword, what are the interactions between the two people...''
\end{quote}

%\textbf{Smart Devices:} Based on the responses from our participants, we found smart device supports (e.g., BeMyEyes, Aira, searching through QR code, object recognition apps) are available but not commonly adopted. Two participants mentioned that they use smart devices for identifying subject (e.g., person, environment, interaction, object) and form (e.g., shape, color, line) information. P15 commented on the use of BeMyEyes in art appreciation practices:

%\begin{quote}
%    ``...I like using Aira or BeMyEyes in everyday tasks, I often use BeMyEyes to ask questions about arts, such as how many people are in the painting, the main colors in the painting...''
%\end{quote}

%Moreover, we found these systems usually lack detailed descriptions, or the person who helps behind the camera does not have expertise in art, this further extend the prior research on RSA (i.e., Remote Sighted Assistance) regarding agents need domain knowledge for specific tasks in addition to this personal knowledge \cite{lee2020emerging}. 
%Beyond using approaches with crowdworkers, our participants also mentioned the past experiences with using image/object recognition apps that automatically identify subjects from visual art. However, we found that they mostly complained about the lack of enough description on contexts of the visual art (e.g., interaction, environment). P7 commented:

% because of these accessibility challenges, they tend to go to the same place. go to the same place also brings emotional values
Because of these aforementioned access challenges with in-person art experiences, our participants mostly \textbf{stick with the same gallery or museum} if they found it to be relatively accessible (as mentioned by six of our participants). Repeated visits to the same venue not only help the staff get familiarized with steps to make the experience accessible, but also bring emotional value to our blind patrons. For example, P11 mentioned:

\begin{quote}
    ``...I usually visit the same art gallery close to my place, the staff there know me pretty well and always reach out to me to ask if I need anything. I feel at home there...''
\end{quote} 

P3 further commented on how visiting the same venue helps consolidate understanding of specific art pieces and reduces uneasy social tensions:

\begin{quote}
    ``...Going to the same art gallery and enjoy(ing) old artwork can help me recall my past memory and enhance [my] understanding of specific art pieces. This makes me feel less discriminated and I could confirm my understanding with docents there...Just like you always get new feelings and inspirations every time you see Mona Lisa...''
\end{quote}

%Particularly, six participants emphasized the preference of \textbf{visiting the \textit{same} art gallery} for perceiving visual arts (P3, P9, P10, P11, P14, P15). The reasons are particularly related to having more accessible art experiences, which include familiarized environments, adaptability to accessibility needs, and less unwanted attention. 

%We also learned that some participants prefer the same art gallery to recall their past memories and understandings of the art. P3 commented:

In summary, our participants' in-person art appreciation experiences leverage many existing non-visual approaches for presenting visual arts, as introduced in the related work (Section \ref{related work}), but with peer- and docent-guided tours being the most commonly adopted. We confirmed on the overall lack of accessibility infrastructures (e.g., tactile graphics) documented in prior work \cite{cavazos2021accessible} and uncovered a common workaround to this problem---familiarizing with and visiting the same art venue over time. We also presented a range of challenges that made their experiences less accessible or made some of these approaches less commonly adopted. One overarching problem is the quality of artwork descriptions, whether it missed descriptions of elements important to artworks or lacked intuitive explanations of visual concepts to blind people. This insight extends prior research on Remote Sighted Assistance (RSA) regarding the need for domain knowledge of specific tasks~\cite{lee2020emerging}. 

%Overall, we learned that our participants have various in-person experiences and preferences with different visual art access methods (e.g., prefer visiting the same art gallery for more accessible experiences). However, we found that none of these current access methods are sufficient for an accessible and enjoyable experience at art galleries and museums. Our findings of the adoption and barriers of existing art access methods provide the art context information related to different access methods in general (e.g., audio descriptions, tactile graphics).

\subsection{Remote Experiences}

Beyond going to art galleries and museums, \textbf{remote art tours} have become increasingly available and popular during COVID-19, and three participants mentioned that they had experiences going to Zoom art tours (P7, P14, P15). They highlighted the benefit of remote tours as a convenient opportunity for blind people to understand visual arts with professional docents, especially during the pandemic. For example, P7 shared his experiences:

\begin{quote}
    ``...I had several experiences with Zoom tours for blind people to enjoy art. I want to say that it has the benefit of accommodating over 300 people at a time, and the docents are usually more professional than normal...because there are usually limited docents that are trained on how to describe visual arts to blind people...''
\end{quote}

However, participants also mentioned the limitations of remote art tours as the lack of engagement and communication, and do not feel as immersive as in-person tours. P15 commented:

\begin{quote}
    ``...there are clear differences compared with in-person experiences. You definitely feel more close to the art if you actually stand in front of it. And you cannot really do much besides listening to the docent from the remote tour...''
\end{quote}

Some participants choose to leverage a combination of remote and in-person access to engage with visual arts, so they can utilize the advantages of both. In particular, P3, P4, and P13 mentioned the benefit of \textbf{searching specific visual arts pieces online} to learn more details about them (e.g., history, artists' background) after visiting an art gallery or museum, as P3 explained:
%Furthermore, we found P3, P4, and P13 leverage \textbf{smart devices to search specific content of visual art} (e.g., history, artists background) as a peripheral approach after coming back from art galleries. P3 further explained the reason behind it:

\begin{quote}
    ``...There is often limited time at art galleries, and the audio descriptions are not often comprehensive enough. I often memorize the visual arts name and come back home and search for it online...''
\end{quote}

\Modified{The increase in remote opportunities for blind people to access visual arts brings attention to the accessibility, engagement, and interactivity of these methods. Our findings provide insights into what blind people enjoy about current remote visual arts accesses (e.g., access to more contextual information, ability to look up and learn more about the art, flexibility in the amount of time), and what challenges they face with them (e.g., less immersive than in-person experiences).} %contextual information on how blind people perceive visual art beyond visiting art galleries in-person and create opportunities to support more interactive and accessible experiences for blind people remotely.

\section{Findings: Perception of Visual Arts Appreciation}
\label{perception}
We now detail how our blind participants perceive visual arts, including their motivations for engaging with visual arts and how they cognitively perceive and encode the meaning and aesthetics of different artworks.

\subsection{Motivation for Engaging with Visual Arts}
\label{motivation}
Our participants have a diverse set of reasons for engaging with visual artworks, including: %We first walk through our blind participants' motivations for appreciating visual arts, which serve as contexts to their experiences and preferences with different visual arts appreciation methods. In total, we present six main motivators: 
1) cultural learning, 2) encouragement, 3) activism, 4) social interactions, 5) relaxation and enjoyments, and 6) nostalgic reasons. 

First, four participants mentioned that appreciating visual arts brings opportunities for \textbf{cultural learning}, as in communicating different individuals' stories to society (P5, P8, P12, P15). These participants are more interested in the background stories and meaning behind visual artworks. P5 explained:

\begin{quote}
    ``...The goal for me to enjoy art is to understand the story and cultural background of different art pieces from different periods. Paintings and pictures are very language independent that you do not need to understand the language that the artists speak to appreciate their paintings...''
\end{quote}

We also found that seven participants consider enjoying visual arts as a way of \textbf{encouragement and inspiration}, especially after their vision loss. P14 mentioned:

\begin{quote}
    ``...My friend brought me to an art gallery right after my vision loss, at that time, I was so depressed that do not even want to step out of my place. Going to the art gallery and getting descriptions from my friend, and discussing different art pieces definitely helped me relieve my stress and discomfort...''
\end{quote}

In particular, three participants perceive visual arts for \textbf{activism purposes}, such as showing sighted people that blind people can also enjoy visual arts at art galleries. P15 highlighted on this:

\begin{quote}
    ``...I really like to go to art galleries with my glasses and cane on, this identifies that I am a blind person and shows to other sighted people that blind people can also appreciate arts. This would increase the awareness of the importance of art accessibility...''
\end{quote}

Moreover, four participants consider going to art galleries as part of  \textbf{social activities} where they could communicate and discuss visual arts with their peers (P3, P4, P11, P14). P3 commented:

\begin{quote}
    ``...Enjoy art can help me with social interaction with other blind people, that is why it is important to understand the detail...We often visit different art galleries together as a social event...''
\end{quote}

We also found that three participants appreciate visual arts for \textbf{relaxation and enjoyment} (P7, P11, P15), as P7 commented:

\begin{quote}
    ``...Exploring art is simple to me, I feel relaxed and connected to the environment while I am in art galleries. It is very similar to watching a movie or playing a game...''
\end{quote}

\begin{figure}[t]
    \centering
    \includegraphics[width=0.5\columnwidth]{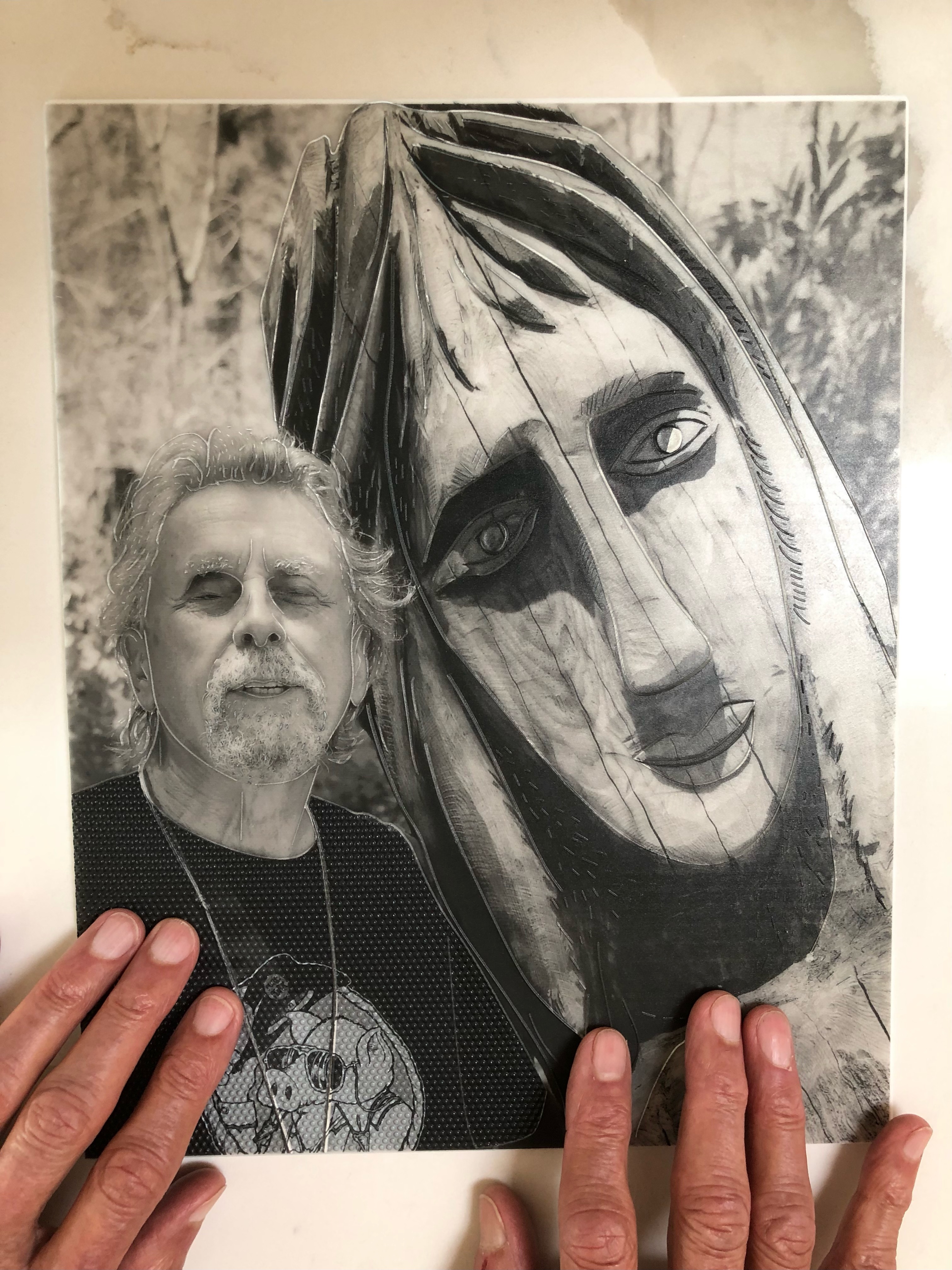}
    \caption{P9 is touching the tactile art piece that was made by an artist to recall his wife who had passed away.}
    \label{fig:tactile}
    \Description{This figure shows a figure that a person is touching the tactile art piece with two hands. There are two faces in the tactile art piece, the left face is a man's face, and the right face is a carving face of a woman.}
\end{figure}

Finally, three participants also engaged with visual arts to recall past memories of people, objects, and events, for \textbf{nostalgic reasons}. For example, P9 spends significant time with a specific piece of visual art in memory of his wife, who had passed away (Figure \ref{fig:tactile}): 

\begin{quote}
    ``...This tactile art was made by an artist to help me recall my wife, who already passed away many years ago. Every time I touched it, I could feel her face and expression, and remember our past memory together...''
\end{quote}

While prior work has listed a range of potential benefits for blind people to non-visually engage with visual arts, we directly consulted blind art patrons' personal perspectives about what aspects of visual arts attract them. Our findings suggest a degree of variance across individuals but point to a set of recurring motivators, many unique to blind people (e.g., activism, nostalgia, and inspiration after losing sight). These motivators serve as contexts for understanding participants' experiences and preferences with different modes of non-visual visual arts presentations.

\subsection{Cognition of Perceiving Visual Arts}
\label{cognition of aesthetic enjoyment}
Here, we present what makes visual arts appreciation aesthetically enjoyable to our participants. First, six participants expressed that they enjoy visual arts through \textbf{imaginations based on existing visual knowledge} and discussing their imaginations with sighted peers (P3, P4, P7, P10, P12, P14). P3 explained:

\begin{quote}
    ``...I used to have vision when I was young, and I currently enjoy art by imagining from the information I know, such as people, activity, and the environment. I then think about what type of color they might use, or the facial expressions, I imagine everything that I am not told. The magic part is confirming my imagination with sighted friends or family members. And it is totally fine if I am wrong, I still like my imagination on how this artwork should be...''
\end{quote}

We found that most participants who experience aesthetic enjoyment through imagination are also people who acquired blindness later on, whereas participants who are congenitally blind have difficulties imagining visual concepts, such as understanding visual references and dimensions. For these participants, the enjoyment instead comes from \textbf{tactile feeling} of the artwork and surrounding environment. P1 commented:

\begin{quote}
    ``...I enjoy touching visual paintings to feel the texture of the painting and even the gallery or museum building to get the atmosphere of the environment. This brings more connections between me and the artwork...''
\end{quote}

Furthermore, some participants also mentioned that enjoyment brought by visual arts is different when accessed through imagination versus tactile renderings. P9 explained:

\begin{quote}
    ``...A tactile version of a drawing or painting does not mean the same thing as the original artwork, it is a reproduction with different meanings and feelings...''
\end{quote}

In sum, our analysis of blind participants' perception of visual arts draws attention to their personal experience (e.g., visual memory), motivation, as well as the presentation mode of visual arts. All these factors influence how aesthetic and artistic visual arts experiences are for blind people.

\section{Findings: Design Considerations for Visual Arts Access Methods}
\label{preference}
In this section, we describe design considerations that influence blind participants' preferences for and adoption of existing visual arts access methods---including peer-guided tours, smart device based access, tactile graphics, audio description, and multimodal presentations. In total, we identified eight design considerations: 1) match comprehension goals with the mode and materials for non-visual access, 2) improve coordination across modalities, 3) balance between flexibility and consistency, 4) enhance objective interpretation, 5) establish shared art vocabulary and grammar, 6) describe visual arts based on individual memory, 7) maintain synchronous feedback, and 8) destigmatize art experiences in social settings.

%1) support comprehension, 2) establish shared art vocabulary and grammar, 3) explore multimodal approaches, 4) avoid personal perceptions, 5) enhance flexibility and ubiquity, 6) maintain feedback as a tight loop, 7) describe visual arts based on individual memory, 8) destigmatize art experiences in social settings.

\subsection{Match Modality Choice with Comprehension Goals}
\label{match modality}

Through our interviews, we learned that the effectiveness of a specific non-visual access method depends on what comprehension goals (i.e., aim to understand different levels of content, such as subject, form, and context \cite{stangl2021going}) a blind patron has toward a visual artwork. In general, participants prefer different access methods to obtain different information. %the importance of considering comprehension and clarity when designing visual art accessing methods for blind people. Overall, we found that participants identify different access methods for different purposes. Among the commonly used approaches, our participants suggest that different access should be used for different comprehension purposes, which affirms prior work on tactile media design \cite{stangl2019defining}. 
For instance, audio descriptions or guided tours are deemed better suited for providing a high-level description of the content (P6, P7, P12, P13), whereas tactile graphics are preferred when participants want direct access to low-level visual references and scenery (P1, P3, P5, P13). P3 explained:

\begin{quote}
    ``...Tactile graphics is hard to tell the big picture of the paintings, but audio could do that easily. However, tactile graphics allows me to feel more details with visual references, such as where things are and some background information as well...''
\end{quote}

In turn, participants commonly want tactile access to
information about the subject (e.g., people, background environment) and form (e.g., shape, line) of a visual art piece, but descriptions (e.g., recorded, from peers or docents) for content (e.g., history of the artist, time period of the work). P4 commented on this:

\begin{quote}
    ``...It is more intuitive to explore the subject and shape of the painting through tactile graphics...I also agree that audio can convey rich content which is supposed to be used to introduce information regarding the history and background of the paintings...''
\end{quote}

Our findings echo past knowledge on the expressive strengths and weaknesses of different modalities~\cite{stangl2019defining,ahmetovic2021touch,morris2018rich}, but uncover how these properties apply to non-visual visual artwork presentations. Because visual artworks often involve more complex and varied types of information compared to other general visual media, adapting the non-visual presentation modality to users' comprehension goals is even more important. 

\subsection{Improve Coordination across Modalities}
\label{improve coordination}
As each access method has its own strengths and weaknesses (as detailed in the previous section \ref{match modality}), many participants would like to combine multiple access methods to enrich their art appreciation experiences. % since different access methods serve different functions (e.g., tactile for easier visual reference, and audio for big picture and richness of the content). 
All the while, having multiple modalities that convey the same content at a time may lead to distraction and confusion. P15 explained:

\begin{quote}
    ``...It is hard to focus on multiple modalities at the same time, unless they serve for different purposes, such as one for guidance and one for specific content. It can distract me if they talk about the same thing...''
\end{quote}

Therefore, having different modalities serve specific functions for which they are suitable and coordinating between these modalities could greatly improve blind patrons' experiences. How to arrange different modalities (e.g., when to play audio descriptions and show tactile representations) should consult the capability and preference of specific users. For example, some blind people are more adept at perceiving tactile signals and may be more comfortable with listening to audio descriptions at the same time as exploring tactile graphics, as P5 explained:

\begin{quote}
    ``...I like having more combinations of methods to explore arts. But I want to say that people who got their blindness at different times have different familiarity with different methods, those who got blind since birth can easily adapt to the tactile approaches, and people like me who did not have many experiences in tactile would like to use audio more. It has to be flexible for us to choose my favorite mappings...''
\end{quote}

\subsection{Balance between Flexibility and Consistency}
\label{balance between flexibility and consistency}
Besides having distinctive experiences with different visual arts access modalities (Section~\ref{improve coordination}), our participants also mentioned a range of other individualized preferences, such as the pace (e.g., fast, slow) and setting (e.g., specific museum location, in-group versus on their own) of visual arts appreciation activities. They commonly expressed the need for more free and flexible access to visual artworks, based on their personal situations. For example, P6 commented on the lack of flexibility in guided tours: 
%From our interview, we uncovered the needs of exploring visual arts freely with different access methods, at different art museums and galleries. We learned that participants prefer approaches that allow for personalized art experiences that enable blind people to explore visual art work at their own pace, such as creating different experiences depending on vision status. They expressed the concerns mostly for guided tours with docents and opportunities of audio descriptions. P6 further explained:

\begin{quote}
    ``...Guided tours for a group of blind people usually follow the same pace depending on the docent, it would be hard for me to further dive into certain artwork in detail if I want to...''
\end{quote}
While participants desire changes in visual arts accesses to tailor to their own needs, a number of them (P3, P4, P13) also shared concerns about having too many changes to keep their experience consistent. In general, our participants value consistency and simplicity in accessibility support, as P4 commented:

\begin{quote}
    ``...It is so common that different galleries have their own ways of obtaining information about arts in a slightly more accessible way to claim as fully accessible, which also creates another layer of barrier for us to adopt their ways while exploring arts in different art galleries, such as some require we search on their websites, some use QR code, and some have pre-recorded audio descriptions through headsets...''
\end{quote}

As too many varied modes for accessing visual arts can be daunting, we learned the importance of balancing between adding options for flexibility and limiting options for consistency. 

\subsection{Enhance Objective Interpretation}
Participants commonly reflected that visual descriptions provided by friends, family, and docents include subjective opinions, making it difficult for them to have individual interpretations of the artwork. This challenge especially interferes with people who consider \textit{imagination} as a way to enjoy the visual art piece (as described in Section \ref{perception}). Instead, they shared the preference for more objective and factual descriptions of the visual arts content, style, and subjects. P2 voiced out this preference:

\begin{quote}
    ``...I prefer to get a perception of the artwork from my own experiences, I do not want to hear personal comments from people, just like this or that painting is so pretty and meaningful, all I need is what color they used, the contours of the lines, and what kinds of objects present in the painting...''
\end{quote}

This preference for fact-based information also applies to services provided by smart devices and remote access methods (as described in Section \ref{current practices and challenges}). P3 explained:

\begin{quote}
    ``...I used to search the detailed content for better understanding of visual arts by searching online through my iPhone, but people usually posted their own feelings and thoughts online regarding certain artwork, this would ruin my whole experiences of arts and destroy my imagination of arts. And there is no such way for me to actually filter it, so I stopped doing that...''
\end{quote}

One workaround many participants mentioned for improving the objectivity of non-visual visual art access is tactile graphics, despite it being more time-consuming and less accessible for some people (as described in Section \ref{current practices and challenges}). This finding confirms the importance of tactile media availability and education (as suggested in previous work such as ~\cite{stangl2019defining,dursin2012information}), but also emphasizes the possibility of making \textit{non-tactile approaches more objective}. 
%Furthermore, we found our participants highly value tactile graphics as they helps blind people make up their own perceptions without interference (P2, P13). All the while, they also commented on the potential improvements for the development of tactile materials: convey more information through tactile graphics and reduce the time cost through tactile graphics (Section \ref{current practices and challenges}).

\subsection{Establish Shared Art Vocabulary and Grammar}
\label{Establish Shared Art Vocabulary and Grammar}
Another challenge our participants highlighted is the lack of commonly understood art vocabularies and expressions across blind people. This challenge is observed for audio descriptions, guided tours, and tactile graphics. 

For description-based methods (e.g., audio descriptions, guided tours), our participants complained about the difficulties for sighted people to describe form information of visual arts (e.g., shape, line, color) in ways they would understand. Describers also do not share a standard for describing visual arts in accessible languages (e.g., vocabulary, grammar, visual references). P15 explained:

\begin{quote}
    ``...I found sighted people always have difficulties explaining color and shape information in detail, which includes the shade of the color, contour of the objects...''
\end{quote}

Furthermore, some participants are also concerned about multi-linguistic and multi-cultural backgrounds affecting their understanding of art vocabulary (P1, P6, P10, P15). For example, P1 shared:

\begin{quote}
    ``...As a blind person from the global south, I often struggle with how different cultural backgrounds affect how we perceive arts. The painting style is different from other types of visual arts as well, this may need a different way to describe it as well...''
\end{quote}

For tactile-based methods, %Similar to audio-based approaches, 
all of our participants encountered barriers with identifying and learning certain expressions in tactile-rendered visual art, highlighting the importance of having common tactile lexicons for arts. P13 commented:

\begin{quote}
    ``...Tactile vocabulary is even less formalized than audio descriptions. This is not surprising at all, as you know, there are not many tactile paintings available for blind people to explore. People create all kinds of tactile versions of paintings, but there is no such standard or guidelines that we should follow...''
\end{quote}

P14 further emphasized the importance of training sessions for people who just lost vision to unify the tactile vocabulary of arts:

\begin{quote}
    ``...It is so important to teach people who just lost their vision how to understand and appreciate visual arts through tactile graphics, it is definitely more direct for blind people. It is not like reading Braille, it is a formalized process of understanding and appreciating arts with location references...''
\end{quote}

This finding is reflective of prior work on descriptions and tactile presentations of visual information~\cite{stangl2019defining, potluri2021examining}, which combined with art literature (e.g., \cite{kleege2016audio}) could provide further guidance on how to standardize accessible vocabularies for blind visual arts patrons.

\subsection{Describe Visual Arts Based on Individual Memory}
Past experience with visual arts is another consideration that our participants often mentioned. Besides the preference for familiar art venues (as described in Section \ref{current practices and challenges}), participants also want their art descriptions to incorporate what they have learned about visual arts in the past, including visual concepts and artwork information. For example, P7 and P12 want the system to provide explanations and links to the visual arts that they know already, as P7 explained:

\begin{quote}
    ``...Every time I see a new painting, it would be useful for the description to refer to certain colors or artistic styles that I have experienced before. This would reduce the time and effort of the descriptions...''
\end{quote}

Moreover, how much and what details to include in the description should also adjust to blind art patrons' memory. For example, it would be beneficial to insert a reminder for blind patrons to quickly know which art pieces they have encountered previously. P6 commented on this issue:

\begin{quote}
    ``...I like to go to the same art gallery to recall my past memory and emphasize my understanding of my favorite paintings. I usually have to ask the surroundings about the painting's name to identify specific artwork. It would be great if I have a system that can tell me when did I interact with this painting and what information did I get last time...''
\end{quote}

Taking past memory into consideration can be particularly beneficial for blind patrons who are in the process of learning visual arts concepts and have difficulties quickly scanning artwork to assess whether they have seen them. 

\subsection{Maintain Synchronous Feedback}
From the interviews, we identified two types of feedback deemed essential by blind patrons (N=$13$) in our study---(1) answers to questions they have about an artwork and (2) comments and discussions of peer patrons (whether sighted or not) about the artwork. Currently, opportunities to clarify details and ask for further details are generally lacking for most access methods (e.g., audio descriptions, and guided tours). For example, P14 explained:

\begin{quote}
    ``...It is hard to verify the audio descriptions by us who are blind, and we cannot ask further questions for audio descriptions if we do not fully understand the content...''
\end{quote}

Similarly, P5 experienced difficulty asking questions during guided tours with docents too:

\begin{quote}
    ``...It can be stressful for me to ask questions to docents and sighted peers when I am in a group of people...''
\end{quote}

Comparatively, listening to other patrons' discussions of artworks is relatively more accessible at art galleries or museums. Many participants expressed their desire to learn about others' interpretation of an artwork, \textbf{after} obtaining an objective understanding of the piece, as P4 commented on this:

\begin{quote}
    ``...After sort of getting to know the artwork and the message behind it, I love to discuss and verify my thoughts with other people, this sometimes can help me add more detailed understandings of the painting...''
\end{quote}

As our interviews point to the importance of communication during visual arts experiences, systems for supporting blind patrons should make efforts to improve and encourage social feedback. 

\subsection{Destigmatize Art Experiences in Social Settings}

While participants highlighted the importance of social communication and feedback for blind visual arts patrons, they also noted that there is social tension when they enjoy visual arts with sighted people. P11 commented:

\begin{quote}
    ``...I do not feel comfortable to keep asking questions during a guided tour with a group of people, it is just too stressful to me to actually enjoy arts...''
\end{quote}

Even if they do not engage in social interactions, blind patrons experience stigma with the way they access visual arts (e.g., touching). Our participants commonly expressed the importance for blind people to present in front of sighted patrons and use tactile methods to explore visual arts, to remove the stigma of touch. P6 explained:

\begin{quote}
    ``...I like to use tactile graphics in front of other sighted folks, I can show them that blind people can also understand and appreciate visual arts through touch. This would reduce the stigma associated with touch and promote the availability of tactile graphics for visual arts...''
\end{quote}

Overall, with the eight design considerations proposed by our participants, our findings further extend prior research on non-visual access to visual information (e.g., \cite{stangl2019defining,kleege2016audio}), with a focus on blind patrons' \textit{unique experiences with visual arts}. With insights on what access methods blind patrons currently adopt, their motivations, challenges, as well as considerations for better non-visual visual arts experiences, we would like to examine how our findings apply broadly to other general blind patrons, including blind people with different backgrounds and vision statuses (e.g., people who are congenitally blind versus who acquired blindness in later life).

\section{Visual Arts Access Preferences Survey}
\label{survey}
Given blind patrons have various demographic backgrounds and vision statuses, we examined how our findings apply broadly to other general blind patrons. We conducted a follow-up survey to examine our interview findings with a larger group of blind people who have experiences in visual arts \cite{anthony2013analyzing,komkaite2019underneath,li2022exploration}. 

\subsection{Method}
\subsubsection{Survey Recruitment}
We recruited survey respondents through the NFB (i.e., National Federation of the Blind) mailing list. To participate in our survey study, the participant needed to be 1) 18 years or older, 2) legally or totally blind, 3) have experiences in visual arts appreciation, and 4) be able to communicate in English. The survey was hosted through Google Forms \cite{GoogleFo84:online}. The survey took about 10 minutes per respondent, and survey respondents who completed the survey were entered into a draw for a \$20 Amazon gift card (1 in every 30). The recruitment and study procedure was approved by the Institutional Review Board (IRB). 

\subsubsection{Survey Questions}
In the survey, we first asked participants screening questions to check whether they meet the previously mentioned recruitment criteria. We then asked for demographic information (e.g., age, gender, details of vision level, onset age of blindness). Next, we asked participants to select up to three preferred methods to access visual arts (according to Section \ref{current practices and challenges}), motivations for engaging in visual arts experience, and sources of aesthetic enjoyment during such experience (according to Section \ref{perception}). Finally, we asked participants to select up to three design considerations that they consider the most important for making their visual arts experiences better (according to Section \ref{preference}). For clarification and ease of understanding, we added examples and explanations of our design considerations in the survey (such as \textit{``have a clear art description standard discussed and agreed with describers, such as what vocabularies to use''} instead of \textit{``establish shared art vocabulary and grammar''}). To potentially reduce biases, we randomized all answers for all questions. We also added an item for survey questions called ``none of these options'' and allowed people to enter their own answers under ``other'' in text if they have any.

\subsubsection{Data and Analysis}
\Modified{Two researchers manually sorted through all 423 responses and the ones that fall under the following categories: 1) incomplete survey (N=$4$), 2) duplicate or malicious entries (e.g., a series of entries with the same age, gender, onset of blindness within a few seconds) (N=$178$), 3) response failing to follow the instruction (e.g., respondents select over three answers for questions we have ``Please choose up to three answers'') (N=$21$). Our data cleaning resulted in 220 valid entries (i.e., 52.0\% of original entries).}

\Modified{Because the goal of this survey is to examine visual arts experiences and preferences learned from our interview with a wider population, our analysis focuses on a descriptive summary of trends across the 220 blind art patrons who answered our survey. For each of the four areas of focus: \textit{motivation}, \textit{preferred access methods}, \textit{sources of enjoyment}, and \textit{design consideration prioritization}, we first summarize the popularity of answer choices among respondents and then compare them across two demographic factors: vision level and the onset of blindness to understand how our findings vary across these factors through inferential statistical tests. Choosing these two factors was determined by our findings from the semi-structured interview (Section \ref{Semi-structured Interview}), such as participants with different blindness onset have different preferences for visual art access methods (Section \ref{preference}). Due to the multiple-response nature of our questions, we treat each answer choice of a question as a binary variable (i.e., if a respondent selected this choice, the value is yes, otherwise no.) and perform chi-squared tests \cite{pearson1900x} to examine the association between the selection of this choice and the two demographic factors of interest (i.e., vision level and the onset of blindness). For testing associations with the onset of blindness, we construct a 2x2 contingency table (\textit{Onset of Blindness} (Congenital, Acquired) x \textit{Choice Selection} (yes, no)), whereas for testing associations with vision level, we construct a 2x3 contingency table (\textit{Vision Level} (Totally blind, Some light perception, Legally blind) x \textit{Choice Selection} (yes, no)). We use a Bonferroni correction \cite{bonferroni1936teoria} to reduce the chance of a Type I error when performing multiple tests \cite{armstrong2014use}.}

\begin{figure*}[t]
    \centering
    \includegraphics[width=1\columnwidth]{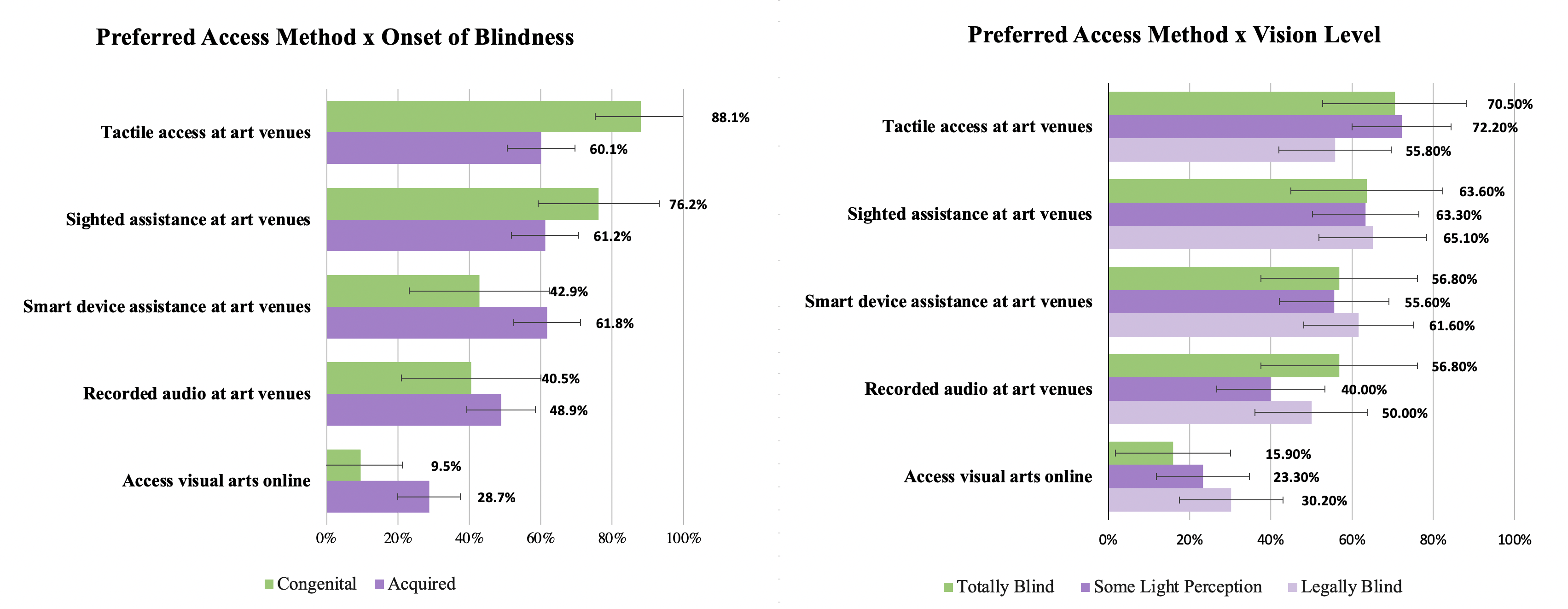}
    \caption{\Modified{Survey respondents' choices of access methods to visual arts appreciation across the onset of blindness (left) and vision level (right).}}
    \label{fig:method}
    \Description{Two bar charts, the left chart shows the preferred access method vs. onset of blindness (congenital, acquired). The y-axis has five different categories (Tactile access at art venues, sighted assistance at art venues, smart device assistance at art venues, recorded audio at art venues, and access visual arts online) and the x-axis has different percentage rates. The right chart shows the preferred access method vs. vision level (totally blind, some light perception, legally blind). The y-axis has five different categories (Tactile access at art venues, sighted assistance at art venues, smart device assistance at art venues, recorded audio at art venues, and access visual arts online) and the x-axis has different percentage rates.}
\end{figure*}

\subsection{Results}
\label{results}
\subsubsection{\Modified{Survey Respondents Demographic Information}}
\Modified{All 220 respondents self-reported to be at least legally blind---86 as ``legally blind'', 90 as ``some light perception'', and 44 as ``totally blind (no light perception).'' 42 of the survey respondents are congenitally blind (acquired blindness since birth). For the remaining 178 participants, the average onset of blindness was $18.96$ years old ($SD = 12.04$). 114 respondents self-identified as male, 102 as female, and 2 preferred not to report gender. The average age of respondents is 34.6 years old (SD = 12.4). All 220 respondents stated having experience engaging with visual arts.}

% \begin{figure}[t]
%     \centering
%     \includegraphics[width=1\columnwidth]{Figures/figure1.png}
%     \caption{Counts of survey respondents' motivations for engaging with visual arts, sources of enjoyment during visual arts appreciation, and preferred access methods.}
%     \label{fig:summary}
%     \Description{to update}
% \end{figure}

\subsubsection{\Modified{Preferred Methods}}
\Modified{In terms of different methods to access visual arts, we found \textit{visiting an art venue with tactile access} ($N = 144$) and \textit{sighted help} ($N = 141$) are still the most preferred access methods to visual arts by the majority of respondents (Table \ref{table:table_sum}), closely followed by \textit{assistance from smart devices at the venue} (N = 128). Only half of the respondents found \textit{pre-recorded audio descriptions} a compelling method to access visual arts at an art venue ($N = 104$), although it is commonly used in practice. \textit{Remote visual arts appreciation} is still only accepted by a smaller group of blind visual art patrons ($N = 54$ out of 220 respondents) (Appendix A, Table \ref{table:table_sum}).}

\Modified{With further analysis of access method preferences across different onsets of blindness (as reported in Table \ref{table:test_method}), we found a significant association ($\alpha = .01$ with Bonferroni correction) between the onset of blindness and the willingness to access visual arts at an art venue through tactile methods ($\chi^2(1, N=220)=10.56, p<.01$). We further dive into different age ranges of blindness onset and we found that our survey respondents with earlier blindness onset tend to prefer to use tactile access more than later blindness onset (Appendix C, Figure \ref{fig:onset_age}). And the percentage of the preference for tactile access becomes similar for people who acquire blindness after 20 years old (Appendix C, Figure \ref{fig:onset_age}).}

% \usepackage{multirow}
\begin{table*}[]
\begin{tabular}{|l|ll|ll|}
\hline
\multirow{2}{*}{\textbf{Preferred Methods}} & \multicolumn{2}{l|}{\textbf{Vision Level}} & \multicolumn{2}{l|}{\textbf{Onset of Blindness}} \\ \cline{2-5} 
                                            & \multicolumn{1}{l|}{$\chi^2$}     & \textit{p}   & \multicolumn{1}{l|}{$\chi^2$}       & \textit{p}       \\ \hline
Recorded audio at art venues                & \multicolumn{1}{l|}{3.77}   & 0.152        & \multicolumn{1}{l|}{0.65}     & 0.419            \\ \hline
Tactile access at art venues                & \multicolumn{1}{l|}{5.84}   & 0.054        & \multicolumn{1}{l|}{10.56}    & \textbf{0.001}   \\ \hline
Sighted assistance at art venues            & \multicolumn{1}{l|}{0.07}   & 0.968        & \multicolumn{1}{l|}{2.68}     & 0.101            \\ \hline
Smart device assistance at art venues       & \multicolumn{1}{l|}{0.71}   & 0.702        & \multicolumn{1}{l|}{4.26}     & 0.039            \\ \hline
Access visual arts online                   & \multicolumn{1}{l|}{8.5}    & 0.014        & \multicolumn{1}{l|}{7.37}     & \textbf{0.007}   \\ \hline
\end{tabular}
\caption{Summary of statistical results from N=220 survey respondents' preferred access methods across the onset of blindness and vision level. Statistical significant results are in bold font ($\alpha=0.01$ with Bonferroni correction).}
\label{table:test_method}
\end{table*}

\Modified{As shown in Figure~\ref{fig:method}, the majority of congenitally blind respondents (88.1\%) prefer having tactile access at art venues, whereas only 60.1\% of respondents who acquired blindness later in life share this preference. We also found a significant association between the onset of blindness and willingness to access visual arts online ($\chi^2(1, N=220)=7.37, p<.01$). We then explore different age ranges of blindness onset and we found our respondents with earlier onset of blindness tend to access visual art online less than those who acquired blindness later (Appendix C, Figure \ref{fig:onset_age}). As shown in Figure~\ref{fig:method}, a much higher percentage of respondents who acquired blindness later on (28.7\%) are willing to access visual art compared to those who are congenitally blind (9.5\%). We found no significant association between the onset of blindness and preferences over sighted assistance, recorded audio, and smart device assistance (Table \ref{table:test_method}).}

\Modified{For analysis of access method preference across vision levels, although we did not find any statistically significant association, we observed interesting trends among our respondents---as shown in Figure~\ref{fig:method}, 30.2\% of legally blind respondents prefer accessing visual arts online, almost double the percentage of totally blind respondents who chose this access method. As Bonferroni correction is a rather conservative approach, future research could further examine this trend, as it could potentially point to limitations in the current design of online visual art access methods that disproportionately influence totally blind art patrons.}

\Modified{Together, these results provide additional evidence to our interview findings that people who acquired blindness at different ages might have different preferences toward accessing methods, especially toward tactile graphics and models as well as remote visual arts activities.}

\begin{figure*}[t]
    \centering
    \includegraphics[width=1\columnwidth]{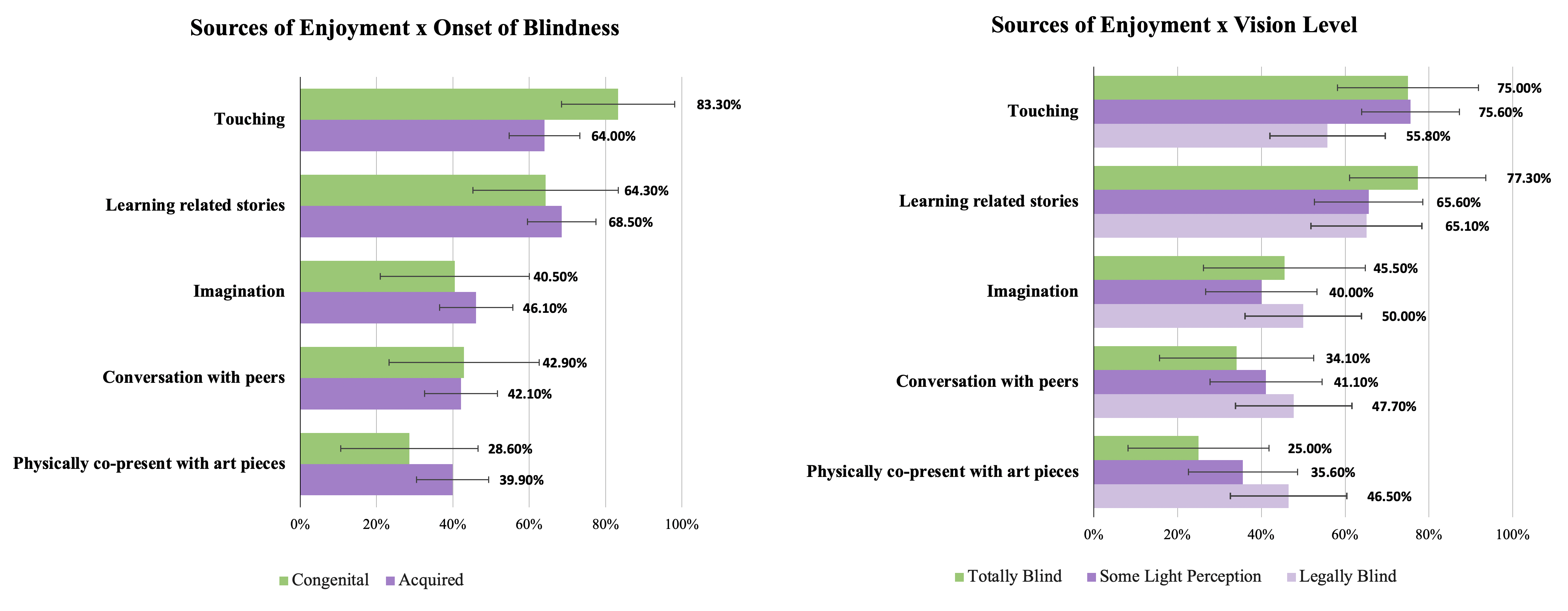}
    \caption{\Modified{Survey respondents' choices of the source of enjoyment for visual arts appreciation across the onset of blindness (left) and vision level (right).}}
    \label{fig:enjoyment}
    \Description{Two bar charts, the left chart shows the sources of enjoyment vs. onset of blindness (congenital, acquired). The y-axis has five different categories (touching, learning related stories, imagination, conversation with peers, and physically co-present with art pieces) and the x-axis has different percentage rates. The right chart shows the sources of enjoyment vs. vision level (totally blind, some light perception, totally blind). The y-axis has five different categories (touching, learning related stories, imagination, conversation with peers, and physically co-present with art pieces) and the x-axis has different percentage rates.}
\end{figure*}

\subsubsection{\Modified{Source of Enjoyment}}

\Modified{From our survey, the majority of survey respondents feel the most aesthetic enjoyment during visual arts appreciation when they get to \textit{learn stories behind the art pieces} ($N=149$) and when \textit{feeling the art piece through touch} ($N=149$) ((Appendix A, Table \ref{table:table_sum})). This result emphasizes the importance of prioritizing both a tactile presentation of visual details and rich stories about the art piece. Meanwhile, almost half of survey respondents also confirmed the joy from \textit{imagining the visuals} ($N = 99$), \textit{conversing with peer art patrons} ($N = 93$), and \textit{physically co-presenting with the artwork} ($N=83$) (Appendix A, Table \ref{table:table_sum})).} 

%Table \ref{table:table_sum} shows respondents' choices (in counts and percentages) of sources of aesthetic enjoyment during visual arts appreciation across vision level and onset of blindness. Overall, we observed that respondents’ choices of sources of enjoyment vary across vision level and onset of blindness. To further quantitatively evaluate these observed trends, we performed statistical tests to examine whether people's sources of enjoyment are associated with their vision level and onset of blindness.

\Modified{In analyzing blind art patrons' source of enjoyment across vision levels and the onset of blindness through a set of chi-squared tests with Bonferroni correction, we found no significant associations. Results of the full set of tests can be found in Appendix B (Table \ref{test_summary}).}

\Modified{A potential area for further exploration is the influence of vision conditions on enjoyment with touching. As shown in Figure~\ref{fig:enjoyment}, a much higher percentage of congenitally blind respondents (83.3\%) in our survey considered touching as a primary source of enjoyment compared to those who acquired blindness (64.0\%), which echos our interview findings (Section ~\ref{perception}). Further, there was also a higher proportion of respondents with less usable vision (75.0\% for totally blind respondents and 75.6\% for those who only have some light perception) who consider touching as their source of enjoyment compared to those with more usable vision (55.8\% for legally blind respondents).}

\begin{figure*}[t]
    \centering
    \includegraphics[width=1\columnwidth]{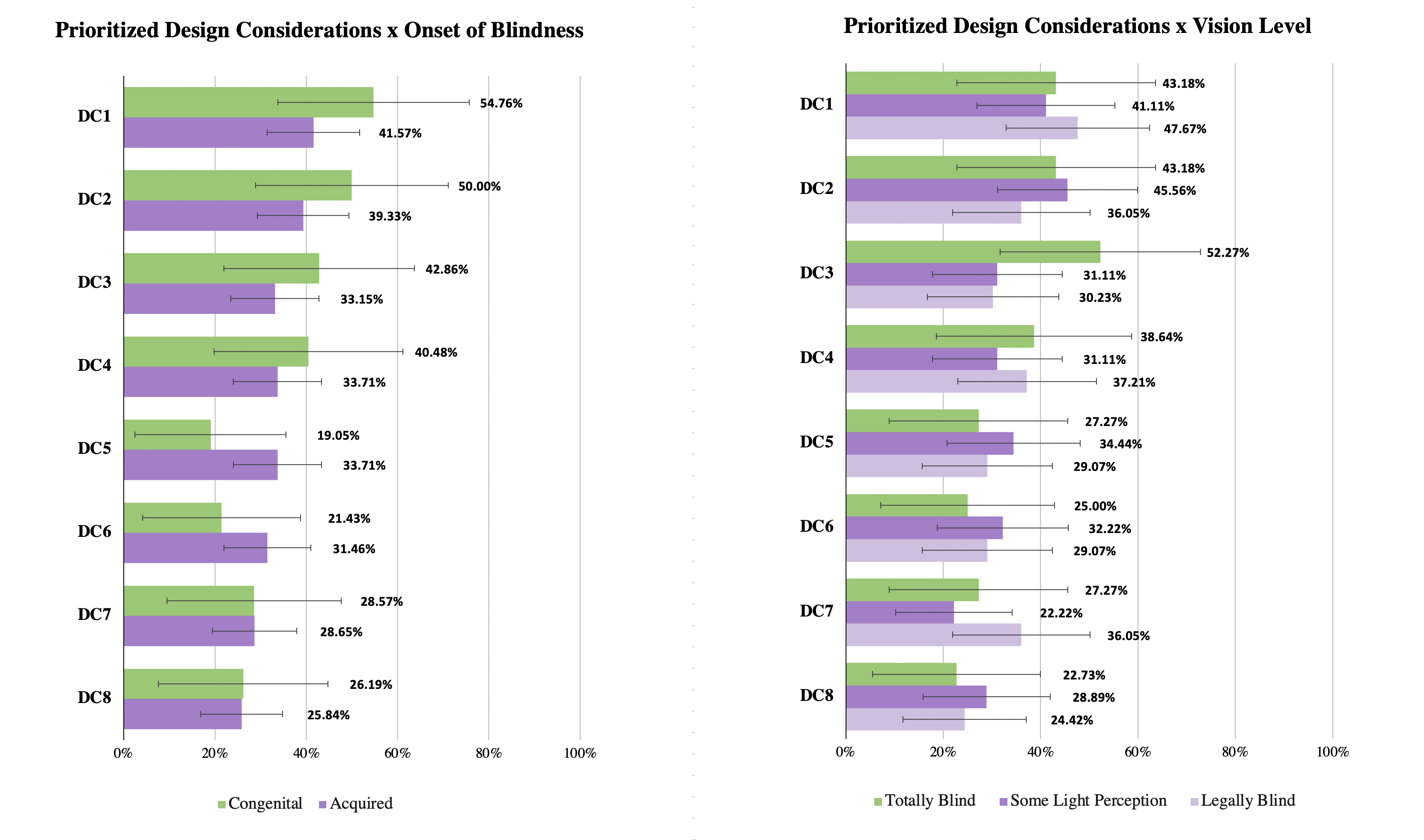}
    \caption{\Modified{Preferences of design considerations to improve the experiences of appreciating visual arts between two groups: onset of blindness and vision level. DC1 = Have the option to access the art piece through multiple modalities (such as tactile and audio) at the same time, DC2 = Have visual details delivered through tactile and background information delivered through audio, DC3 = Have a clear art description standard discussed and agreed with describers-such as what vocabularies to use, DC4 = Have flexibility over the pace \& location \& time of art appreciation experience, DC5 = Have a social setting where you feel more comfortable and less stigmatized to enjoy visual art, DC6 = Have the ability to filter out subjective interpretations in the art description-such as those from sighted peers or docents, DC7 = Have art descriptions adapted to your experience and knowledge-such as skipping information that you already knew, DC8 = Have the ability to acquire answers from someone you trust whenever you have a question.}}
    \label{fig:designconsideration}
    \Description{Two bar charts, the left chart shows the prioritized design considerations vs. onset of blindness (congenital, acquired). The y-axis has eight different categories (DC1 - DC8) and the x-axis has different percentage rates. The right chart shows the prioritized design considerations vs. vision level (totally blind, some light perception, totally blind). The y-axis has eight different categories (DC1 - DC8) and the x-axis has different percentage rates.}
\end{figure*}

\subsubsection{\Modified{Motivation}}
\Modified{Our survey results suggest that four motivations for engaging in visual arts experience that we learned from interviews also seem to be common across 220 survey respondents, including \textit{learning cultural knowledge and story} ($N=143$), \textit{relaxation and enjoyment} ($N=116$), \textit{social interaction and involvement} ($N=112$), as well as \textit{encouragement and inspiration} ($N=109$) (full results shown in Appendix A). While not as common, a quarter of respondents ($N=53$) also access visual arts for \textit{nostalgic reasons}, and 39 for \textit{achieving activism-related goals}.}

\Modified{Our chi-squared tests with Bonferroni correction suggest no significant association between motivation to engage in visual arts appreciation and visual conditions. Results of the full set of tests can be found in Appendix B (Table \ref{test_summary}).}

%We found 81.0\% of respondents who are congenitally blind are motivated to learn the culture, knowledge, and story of different art pieces, and the number of people who acquired blindness later in life was only 61.2\%. We further ran Fisher’s exact tests \cite{fisher1992statistical} on touching and we found the p-value equals to 0.019, which is not statistically significant with Bonferroni correction ($\alpha=0.0083$) \cite{bonferroni1936teoria}. We also conducted Fisher's exact tests for all other answer choices from the \textit{motivation} question across respondents' onset of blindness. We did not identify statistical significance which means there are no statistical differences across sources of enjoyment for the onset of blindness.}

\subsubsection{\Modified{Design Considerations}}

\Modified{All design considerations we learned from the interview received endorsement from at least one fourth of the respondents, as shown in Appendix A (Table \ref{table:table_sum}). In particular, the most popular consideration is to \textit{have the option of accessing art pieces through multiple modalities at the same time} ($N = 97$). 91 respondents further indicated that \textit{visual details should be delivered through tactile, and background information should be delivered through audio}. Moreover, 77 respondents chose that \textit{having a clear standard of vocabularies for art descriptions agreed between the describer and receiver} is important for improving their art appreciation experience, while the \textit{flexibility over art appreciation pace, location and time} are also important to many respondents ($N = 77$) (Appendix A, Table \ref{table:table_sum}).} 

\Modified{In analyzing associations between the perceived importance of design considerations and vision conditions, our chi-squared tests with Bonferroni correction suggest no significant results. Results of the full set of tests can be found in Appendix B (Table \ref{test_summary}). However, a potential future area of exploration is any social and psychological challenges brought by acquiring blindness later in life, as we observed that overall more survey respondents who acquired blindness (33.7\%) treated social stigmatization as a priority among all design considerations we listed, compared to those who are congenitally blind (19.05\%) (as shown in Figure~\ref{fig:designconsideration})}. %To verify, we then conducted Fisher's exact tests \cite{fisher1992statistical} on ``Have a social setting where you feel more comfortable and less stigmatized to enjoy visual art'' across different onset of blindness. We found the difference was not statistically significant ($p=0.0936$). We also found all other results, according to onset of vision impairment, revealed that both groups have similar preferences on design considerations (Figure \ref{fig:designconsideration}).}

\section{Discussion}

\label{discussion}
\Modified{In this section, we discuss the summary of our interview and survey studies, implications for context-based visual art experiences, and focused technological opportunities. We first discuss how our interview findings and survey results connect with past literature and corresponding design implications for future customization of access support for blind patrons (Section \ref{Summary of Design Implications for Blind Art Appreciation}). We then discuss in-depth future design directions relevant to \textbf{visual art experiences under different contexts} (e.g., Social Settings and Activism (Section \ref{Art Appreciation in Social Settings and Activism}), Universal Design for Visual Art Ecosystem (Section \ref{Universal Design for Visual Arts Ecosystem}), and Remote Art Access Experiences Support (Section \ref{Remote Art Access Experiences Support})). Finally, we discuss \textbf{technological opportunities to enhance visual art access methods} for blind patrons (Audio Description Design beyond Accuracy (Section \ref{Audio Description Design Beyond Accuracy}), Context-based Automatic Image Recognition (Section \ref{Context-based Automatic Image Recognition}), and Multimodal Approach towards Different Vision Conditions (Section \ref{multimodal Approach towards Blind Onset})).} %(e.g., remote art access experiences support, universal design for art gallery and museum, art appreciation in social settings, context-based automatic image recognition, and audio description design beyond clarity). 

\subsection{Summary of Design Implications for Blind Art Appreciation}
\label{Summary of Design Implications for Blind Art Appreciation}
Through 15 in-depth interviews and a large-scale survey, we learned about blind people's diverse interests and experiences with visual arts. The survey results confirmed a set of common practices, challenges, motivations, as well as considerations of blind visual arts patrons that we learned from the interviews (Section \ref{results}). While visiting art venues with tactile graphics or with sighted peers are the most popular access methods for blind patrons, we see promises in technology-based support to transform these methods, especially through better coordinated multimodal experiences and personalized design. In particular, both the interviews and survey suggest that personal factors, including vision level, the onset of blindness, motivation, comprehension goals, past experiences with visual arts, and non-visual presentation modalities all contribute to individuals' unique needs for personalized visual arts access solutions, which should not box-in anyone with a vision impairment as one-size-fits-all.
Our findings echo with existing research that identified people who acquired blindness at different stages of life as having various ``blind'' perceptions toward spatial and temporal information~\cite{landgraf2013see}. Therefore, we suggest future research on visual arts accessibility to consider co-design, participatory design, and longitudinal involvement with blind people at different stages (e.g., different vision level, different years of visual experiences), that supports learning, technology adoption, and social factors in their life~\cite{chick2017co,azenkot2016enabling}.

We also learned a detailed set of challenges and considerations with numerous aspects of non-visual visual arts appreciation, centering on access modalities, visual concept comprehension, interpretation freedom, convenience, and social tensions. These considerations call to past work on multimodality (e.g.,~\cite{candlin2003blindness,li2022freedom,kress2009multimodality}), non-visual presentations of visual information (e.g., ~\cite{li2021non,siu2022supporting}), and social considerations of assistive technologies (e.g., \cite{shinohara2011shadow,li2021choose,li2022feels}), but specifically provide guidance for visual arts presentations. In particular, these findings provide novel insights on what blind patrons want to learn about specific elements of visual arts (e.g., form, shape, or content), which of them should be presented in what modalities (e.g., haptics, audio, level of interactivity) (Section \ref{match modality}), persisting challenges with these modalities, and what types of social and interpretation goals are involved (Section \ref{current practices and challenges}). %Given that prior research explored and suggested multisensory design for blind people \cite{candlin2003blindness}, we found that it is important to match the modality choice with comprehension goals (Section \ref{match modality}) and improve coordination across modalities (e.g., not having multiple modalities convey the same content at the same time) (Section \ref{improve coordination}). We suggest future researchers consider the comprehension and cognition of different elements of vision arts (e.g., form, shape, content) with different modalities (Section \ref{preference}) \cite{kress2009multimodality}.
Overall, we advocate for future research to incorporate considerations identified in our study when designing for blind people's access to visual arts. Below, we further suggest possible research directions.

\subsection{\Modified{Art Appreciation in Social Settings and Activism}}
\label{Art Appreciation in Social Settings and Activism}
\Modified{As suggested in our findings, visual arts appreciation can be a social activity for blind people (e.g., discussing paintings in detail with their friends, attending visual arts venue tours, and activism). Future access methods should go beyond just providing information about virtual arts and take social interactions into consideration, as suggested for general accessibility support~\cite{shinohara2011shadow}. Beyond building applications and social infrastructures that connect blind patrons with peers or docents with expertise, future research could also consider opportunities and challenges for utilizing Human-Robot Interaction to mitigate the stress of people who feel anxious about social interactions~(e.g., \cite{tsui2015accessible}).}

\Modified{As some of our participants prefer going to art galleries with their \textit{blind} peers, another future research area is to explore dynamics and collaboration techniques \cite{bonani2018my,mendes2020collaborative} for blind patron groups, and investigate associated design opportunities. Furthermore, we also learned that our participants perceive visual arts for activism purposes (Section \ref{motivation}). According to the prior definition, ``art activism'' has shown the ability of art to function as the medium for social activism \cite{reed2019art}. Despite the legal right to activism and protection against discrimination under the Rehabilitation Act (Section 504) \cite{RehabAct1973}, in-person activism poses additional barriers for people with disabilities as found by Li et al. \cite{li2018slacktivists}. In addition to the institutional barriers against activism, people with disabilities have to navigate additional physical and environmental accessibility challenges to effectively collectivize and protest against (disability) injustice. Thus, more should be explored for blind activists in art museums and galleries \cite{romeo2018access}, such as leveraging mobile AR applications to support blind activists' interactions with the space and reduce physical barriers with other people \cite{herskovitz2020making}.}

\subsection{\Modified{Universal Design for Visual Arts Ecosystem}}
\label{Universal Design for Visual Arts Ecosystem}
\Modified{Accessibility of visual arts involves not only engaging with the art piece itself, but also navigating to and within art venues, locating art pieces, manipulating related technologies (e.g., QR code, website search, audio headsets)~\cite{alghamdi2013indoor,sato2017navcog3,romeo2022maps}, interacting with others, and so on. Exploring and learning all these modules while figuring out what a visual art piece is about non-visually can be daunting. Our participants expressed concerns over inconsistent, new accessibility attempts across different art venues (Section \ref{balance between flexibility and consistency}), which was one reason they preferred revisiting the same gallery for appreciating visual arts (Section \ref{current practices and challenges}). %In our findings, we demonstrated the preference of supporting flexibility and ubiquity for art accessing methods (Section \ref{balance between flexibility and consistency}). We learned that there is a high variance of art access methods for different art venues (e.g., QR code, searching through websites, pre-recorded audio headsets). Beyond just gaining access to specific visual art piece, appreciating arts also require the ability to navigate in the art gallery and the ability to switch between different art pieces. Existing research has explored various approaches of using WiFi, RFID, and QR code for indoor navigation of blind people (e.g., \cite{alghamdi2013indoor,sato2017navcog3}). 
We therefore suggest future research on this topic to consider the whole ecosystem of visual arts appreciation experiences with a Universal Design lens~\cite{iwarsson2003accessibility,persson2015universal}. For example, more effort should be put into providing standardized, intuitive control and accessibility support to reduce the effort of switching between devices, adjusting access methods or languages, and learning unfamiliar processes.}

\subsection{Remote Art Access Experiences Support}
\label{Remote Art Access Experiences Support}
From our interviews, we learned many acknowledged benefits of online art tours (e.g., allows large group attendance, more flexibility, less social tension) (Section \ref{current practices and challenges}). However, there are also critical limitations with existing remote tour tools, mainly the lacked sense of engagement of remote blind attendees and communication barriers with peers and docents. During COVID-19, increasing research has explored how to support blind people with online conferencing tools \cite{ellis2021smartphones,leporini2021distance} and associated tasks (e.g., raising/unraising hands, using the chat, checking microphone and camera status) (\cite{leporini2021distance}). We suggest future research on accessible remote visual arts engagement to consult guidance from this work, and further develop support for social interactions of blind patrons for them to feel immersed \Modified{remotely. For example, future research could explore external hardware pieces that can adaptively and automatically provide tactile experiences, such as Force Feedback Tablet (F2T) \cite{pissaloux2022new,gay2018towards} and Graphical Braille Display \cite{gyoshev2018exploiting}), which could potentially be adopted for remote experiences for blind people.} Moreover, future remote art accessing tools could also include features for customizing non-visual presentations based on blind users' comprehension goals, preferred art vocabularies, visual experiences, and so on (as suggested in Section \ref{preference}). Finally, future tools should provide enhancement to visual arts experiences with opportunities through online content \cite{batanero2021improving} or previous experiences \cite{li2019fmt}. %How to include and design these experiences involve complex considerations of user needs, accessibility, and technology feasibility, keeping this problem area wide and challenging. %Here we offer a set of research questions for future research to consider: 1) how should conferencing tools be used to support blind visual arts experiences in a social manner? 2) how should conferencing tools provide customized presentation and comprehension for virtual art experiences? 3) how should conferencing tools provide enhancement to the art experiences with opportunities through online content?

%Beyond interaction methods, we also identified that culture and languages may affect how people perceive visual arts (P1), which was one reason some participants preferred revisiting the same gallery for appreciating visual arts. Therefore, our finding of ``choosing the same art gallery or museum'' brings more opportunities for future research to explore universal design \cite{iwarsson2003accessibility,persson2015universal} on supporting more opportunities for blind people to access visual arts in a space easily by given their familiarized methods. 

\subsection{\Modified{Audio Description Design beyond Accuracy}}
\label{Audio Description Design Beyond Accuracy}
% artistic experience
% support interactive opportunities for personal understanding
% toggle between objective information and subjetive opinions
\Modified{Our findings uncover a set of criteria for audio description design beyond clarity and accuracy. One important criterion is to encourage an artistic experience of visual arts descriptions, by providing information that supports blind patrons' visual imaginations. For example, the descriptions should use references and metaphors familiar and aesthetic to blind patrons in explaining visual concepts. We also found that blind patrons enjoy interactive opportunities while listening to audio descriptions. We propose audio description interfaces to support users with more agency and control, such as by allowing them to select specific pieces of information to listen to, save progress, and control speed as well as the level of details to match personal exploration paces. Importantly, blind patrons should be able to toggle between objective descriptions and subjective opinions. With these potentially complex interaction steps,
there remain opportunities to involve automatic speech recognition \cite{azenkot2013exploring} to make blind users' control less interrupting to their art experiences. Beyond using speech input, this also brings opportunities for researchers to leverage motion tracking to support blind people by leveraging different gestures to interact with audio interfaces more naturally and easily in a museum context \cite{dim2014designing,li2017braillesketch,sun2021teethtap,sun2021thumbtrak}.} %save and extract the content from professional docents in the tour and further use it for audio descriptions for blind people to appreciate arts independently.

\subsection{Context-based Automatic Image Recognition}
\label{Context-based Automatic Image Recognition}
Despite previous efforts in generating automatic image captions for visual artworks (e.g., ~\cite{ahmetovic2021touch,morris2018rich}), our participants still commonly experience frustrations using computer vision tools to acquire information about visual arts---often, visual details and contexts essential to these artworks are missing (Section \ref{current practices and challenges}). We believe further advancement is needed to consider what information of a visual art piece blind patrons actually care about, for which our findings provide hints---they are generally interested in the content, subject, and form of an artwork, but this interest varies with personal goals, experiences, knowledge, and cultures (Section \ref{preference}). How to allow blind patrons to customize what information is included in visual arts captions based on personal contexts (e.g., provide several alternative versions of such captions, provide information based on questions) still remains an open question.
%It is also important to investigate interactions between the system and blind people and create new interations that allows blind people to adjust how detail they would like to receive certain information about arts.

\subsection{\Modified{Multimodal Approach towards Different Vision Conditions}}
\label{multimodal Approach towards Blind Onset}
\Modified{In our research, we uncovered the diversity of perceptions, practices, and challenges from our participants with different blindness onset and progression. Visual arts access for blind people would require more of a personalized approach to meet blind patrons' preferences and needs that differ across vision level, past visual memory, as well as other co-existing health conditions. To encourage and prepare blind people with necessary knowledge and resources for visual arts experiences (e.g., art vocabulary and grammar in both audio and tactile forms), we suggest future research to build multimodal learning tools (e.g., \cite{coughlan2022non}) (Section \ref{Establish Shared Art Vocabulary and Grammar}). Providing arts education in different modalities would reduce the effort of learning and provide more immersive experiences for blind people with different visual conditions. Existing art museums and galleries rarely provide art experiences for blind patrons in a non-visual modality, let alone a set of modalities (e.g., tactile graphics and audio descriptions~\cite{romeo2018simplification,cantoni2018art})). These art venues should explore ways to provide different modalities, and more research is needed to explore various approaches to create low-cost multimodal systems \cite{duy2022semi} with blind patrons \cite{reichinger2018designing} in art museums. For example, Dao et al. \cite{duy2022semi} explored semi-automatic contour “gist” creations for museum painting tactile exploration to reduce the labor cost and high effort. Beyond leveraging multimodal approaches to enhance the richness of information for visual arts, future research should also explore gesture sets for different modalities~\cite{romeo2022inclusive,reichinger2018pictures,reichinger2016gesture} in a customizable fashion to enable accessible interactive visual art accessing experiences.}

\section{Limitation and Future work}
In our work, we chose to focus on exploring visual arts appreciation of blind people only (i.e., who are legally blind or totally blind). People with other low vision conditions (e.g., blurred vision, loss of peripheral vision, central vision loss \cite{LowVisio59:online}) may have different practices, perceptions, challenges, and preferences of visual arts appreciation. We recommend future research to critically consider how our findings may apply to a wider range of people with low vision. We also call for more work to explore how people with low vision engage with visual arts differently from what we have learned. Our initial identification of key challenges and practices of blind patrons was limited to 15 interviews. On a topic as subjective and complex as artistic experiences, there could be more individual differences within the blind community, which we attempted to count through the survey, but may still be missed due to the nature of the survey methodology. We encourage future work to not treat insights from our study as definitive but continuously explore diverse perspectives. \Modified{Moreover, we chose to leverage interview and survey to explore blind people's experiences, perceptions, and preferences of visual art access. Future research could also conduct contextual inquiries to understand specific practices or challenges in-depth (e.g., \cite{kianpisheh2019face}).} Last, to prioritize the accessibility of our survey, we chose to use a platform that provides less control over respondents' behavior, and thus could not provide warnings when they answered a question not as directed (e.g., chose more than three answers to certain questions). We hope to see more accessibility advancements in professional survey tools to strengthen exploratory accessibility studies as the one in this paper. 
%In our survey, we asked our participants to ``choose up to three answers'' for some questions, which cannot identify the ranks among these answers. 

\section{Conclusion}
In this paper, we describe the findings of an interview study involving 15 blind visual arts patrons and a follow-up survey study (N=220) to understand their current and potential future use of technology for art appreciation experiences. We highlight the importance of aesthetic enjoyment in non-visual visual arts presentations and provide insights on why and how blind patrons experience the sense of appreciation from visual arts (e.g., imagination, tactile senses). We provide eight design considerations for future research and development efforts to consult in advancing visual arts accessing methods for the blind. Our work extends existing research on non-visual presentations of visual information and accessibility for blind people, by contributing design insight specifically related to visual arts. In turn, we propose a list of future research directions (e.g., remote art experiences support, \Modified{multimodal approach towards different vision conditions}) to support blind people with more independent and interdependent access to visual arts, leveraging multiple modalities, machine learning, and personalization technologies.

\begin{acks}
This project is supported by CMU HCII, UW CREATE, and the NSF/CRA CIFellows program.
\end{acks}

\bibliographystyle{ACM-Reference-Format}
\bibliography{main}


\newpage
\appendix
\onecolumn
\section{Survey Response Summary}

% \usepackage{multirow}
% \usepackage{multirow}

% \usepackage{multirow}
\begin{table}[h]
\begin{tabular}{|l|llllll|}
\hline
\multirow{3}{*}{}                     & \multicolumn{1}{l|}{\multirow{2}{*}{\textbf{Total}}} & \multicolumn{3}{l|}{\textbf{Vision Level}}                                                             & \multicolumn{2}{l|}{\textbf{Onset of Blindness}} \\ \cline{3-7} 
                                      & \multicolumn{1}{l|}{}                                & \multicolumn{1}{l|}{TB}          & \multicolumn{1}{l|}{LP}          & \multicolumn{1}{l|}{LB}          & \multicolumn{1}{l|}{C}            & A            \\ \cline{2-7} 
                                      & \multicolumn{1}{l|}{220 (100.0\%)}                   & \multicolumn{1}{l|}{44 (20.0\%)} & \multicolumn{1}{l|}{90 (40.9\%)} & \multicolumn{1}{l|}{86 (39.1\%)} & \multicolumn{1}{l|}{42 (19.1\%)}  & 178 (80.9\%) \\ \hline
\textbf{Preferred Methods}            & \multicolumn{6}{l|}{}                                                                                                                                                                                            \\ \hline
Tactile access at art venues          & \multicolumn{1}{l|}{144 (65.5\%)}                    & \multicolumn{1}{l|}{31 (70.5\%)} & \multicolumn{1}{l|}{65 (72.2\%)} & \multicolumn{1}{l|}{48 (55.8\%)} & \multicolumn{1}{l|}{37 (88.1\%)}  & 107 (60.1\%) \\ \hline
Sighted assistance at art venues      & \multicolumn{1}{l|}{141 (64.1\%)}                    & \multicolumn{1}{l|}{28 (63.6\%)} & \multicolumn{1}{l|}{57 (63.3\%)} & \multicolumn{1}{l|}{56 (65.1\%)} & \multicolumn{1}{l|}{32 (76.2\%)}  & 109 (61.2\%) \\ \hline
Smart device assistance at art venues & \multicolumn{1}{l|}{128 (58.2\%)}                    & \multicolumn{1}{l|}{25 (56.8\%)} & \multicolumn{1}{l|}{50 (55.6\%)} & \multicolumn{1}{l|}{53 (61.6\%)} & \multicolumn{1}{l|}{18 (42.9\%)}  & 110 (61.8\%) \\ \hline
Recorded audio at art venues          & \multicolumn{1}{l|}{104 (47.3\%)}                    & \multicolumn{1}{l|}{25 (56.8\%)} & \multicolumn{1}{l|}{36 (40.0\%)} & \multicolumn{1}{l|}{43 (50.0\%)} & \multicolumn{1}{l|}{17 (40.5\%)}  & 87 (48.9\%)  \\ \hline
Access visual arts online             & \multicolumn{1}{l|}{54 (24.6\%)}                     & \multicolumn{1}{l|}{7 (15.9\%)}  & \multicolumn{1}{l|}{21 (23.3\%)} & \multicolumn{1}{l|}{26 (30.2\%)} & \multicolumn{1}{l|}{3 (9.5\%)}    & 51 (28.7\%)  \\ \hline
\textbf{Source of Enjoyment}          & \multicolumn{6}{l|}{}                                                                                                                                                                                            \\ \hline
Touching                              & \multicolumn{1}{l|}{149 (67.7\%)}                    & \multicolumn{1}{l|}{33 (75.0\%)} & \multicolumn{1}{l|}{68 (75.6\%)} & \multicolumn{1}{l|}{48 (55.8\%)} & \multicolumn{1}{l|}{35 (83.3\%)}  & 114 (64.0\%) \\ \hline
Learning related stories              & \multicolumn{1}{l|}{149 (67.7\%)}                    & \multicolumn{1}{l|}{34 (77.3\%)} & \multicolumn{1}{l|}{59 (65.6\%)} & \multicolumn{1}{l|}{56 (65.1\%)} & \multicolumn{1}{l|}{27 (64.3\%)}  & 122 (68.5\%) \\ \hline
Imagination                           & \multicolumn{1}{l|}{99 (45.0\%)}                     & \multicolumn{1}{l|}{20 (45.5\%)} & \multicolumn{1}{l|}{36 (40.0\%)} & \multicolumn{1}{l|}{43 (50.0\%)} & \multicolumn{1}{l|}{17 (40.5\%)}  & 82 (46.1\%)  \\ \hline
Conversation with peers               & \multicolumn{1}{l|}{93 (42.3\%)}                     & \multicolumn{1}{l|}{15 (34.1\%)} & \multicolumn{1}{l|}{37 (41.1\%)} & \multicolumn{1}{l|}{41(47.7\%)}  & \multicolumn{1}{l|}{18 (42.9\%)}  & 75 (42.1\%)  \\ \hline
Physically co-present with art pieces & \multicolumn{1}{l|}{83 (37.7\%)}                     & \multicolumn{1}{l|}{11 (25.0\%)} & \multicolumn{1}{l|}{32 (35.6\%)} & \multicolumn{1}{l|}{40 (46.5\%)} & \multicolumn{1}{l|}{12 (28.6\%)}  & 71 (39.9\%)  \\ \hline
\textbf{Motivation}                   & \multicolumn{1}{l|}{}                                & \multicolumn{1}{l|}{}            & \multicolumn{1}{l|}{}            & \multicolumn{1}{l|}{}            & \multicolumn{1}{l|}{}             &              \\ \hline
Learning                              & \multicolumn{1}{l|}{143 (65.0\%)}                    & \multicolumn{1}{l|}{30 (68.2\%)} & \multicolumn{1}{l|}{62 (68.9\%)} & \multicolumn{1}{l|}{51 (59.3\%)} & \multicolumn{1}{l|}{34 (81.0\%)}  & 109 (61.2\%) \\ \hline
Encouragement                         & \multicolumn{1}{l|}{109 (49.6\%)}                    & \multicolumn{1}{l|}{14 (31.8\%)} & \multicolumn{1}{l|}{49 (54.4\%)} & \multicolumn{1}{l|}{46 (53.5\%)} & \multicolumn{1}{l|}{18 (42.9\%)}  & 91(51.1\%)   \\ \hline
Activism                              & \multicolumn{1}{l|}{39 (17.7\%)}                     & \multicolumn{1}{l|}{7 (15.9\%)}  & \multicolumn{1}{l|}{15 (16.7\%)} & \multicolumn{1}{l|}{17 (19.8\%)} & \multicolumn{1}{l|}{7 (16.7\%)}   & 32 (18.0\%)  \\ \hline
Social                                & \multicolumn{1}{l|}{112 (50.9\%)}                    & \multicolumn{1}{l|}{23 (52.3\%)} & \multicolumn{1}{l|}{40 (44.4\%)} & \multicolumn{1}{l|}{49 (57.0\%)} & \multicolumn{1}{l|}{26 (61.9\%)}  & 86 (48.3\%)  \\ \hline
Relaxation                            & \multicolumn{1}{l|}{116 (52.7\%)}                    & \multicolumn{1}{l|}{25 (56.8\%)} & \multicolumn{1}{l|}{51 (56.7\%)} & \multicolumn{1}{l|}{40 (46.5\%)} & \multicolumn{1}{l|}{22 (52.4\%)}  & 94 (52.8\%)  \\ \hline
Nostalgic                             & \multicolumn{1}{l|}{53 (24.1\%)}                     & \multicolumn{1}{l|}{11 (25.0\%)} & \multicolumn{1}{l|}{14 (15.6\%)} & \multicolumn{1}{l|}{28 (32.6\%)} & \multicolumn{1}{l|}{5 (11.9\%)}   & 48 (27.0\%)  \\ \hline
\textbf{Design Considerations}        & \multicolumn{1}{l|}{}                                & \multicolumn{1}{l|}{}            & \multicolumn{1}{l|}{}            & \multicolumn{1}{l|}{}            & \multicolumn{1}{l|}{}             &              \\ \hline
DC1                                   & \multicolumn{1}{l|}{97 (44.1\%)}                     & \multicolumn{1}{l|}{19 (43.2\%)} & \multicolumn{1}{l|}{37 (41.1\%)} & \multicolumn{1}{l|}{41 (47.7\%)} & \multicolumn{1}{l|}{23 (54.8\%)}  & 74 (41.6\%)  \\ \hline
DC2                                   & \multicolumn{1}{l|}{91 (41.4\%)}                     & \multicolumn{1}{l|}{19 (43.2\%)} & \multicolumn{1}{l|}{41 (45.6\%)} & \multicolumn{1}{l|}{31 (36.0\%)} & \multicolumn{1}{l|}{21 (50.0\%)}  & 70 (39.3\%)  \\ \hline
DC3                                   & \multicolumn{1}{l|}{77 (35.0\%)}                     & \multicolumn{1}{l|}{23 (52.3\%)} & \multicolumn{1}{l|}{28 (31.1\%)} & \multicolumn{1}{l|}{26 (30.2\%)} & \multicolumn{1}{l|}{18 (42.9\%)}  & 59 (33.1\%)  \\ \hline
DC4                                   & \multicolumn{1}{l|}{77 (35.0\%)}                     & \multicolumn{1}{l|}{17 (38.6\%)} & \multicolumn{1}{l|}{28 (31.1\%)} & \multicolumn{1}{l|}{32 (37.2\%)} & \multicolumn{1}{l|}{17 (40.5\%)}  & 60 (33.7\%)  \\ \hline
DC5                                   & \multicolumn{1}{l|}{68 (30.9\%)}                     & \multicolumn{1}{l|}{12 (27.3\%)} & \multicolumn{1}{l|}{31 (34.4\%)} & \multicolumn{1}{l|}{25 (29.1\%)} & \multicolumn{1}{l|}{8 (19.0\%)}   & 60 (33.7\%)  \\ \hline
DC6                                   & \multicolumn{1}{l|}{65 (29.5\%)}                     & \multicolumn{1}{l|}{11 (25.0\%)} & \multicolumn{1}{l|}{29 (32.2\%)} & \multicolumn{1}{l|}{25 (29.1\%)} & \multicolumn{1}{l|}{9 (21.4\%)}   & 56 (31.5\%)  \\ \hline
DC7                                   & \multicolumn{1}{l|}{63 (28.6\%)}                     & \multicolumn{1}{l|}{12 (27.3\%)} & \multicolumn{1}{l|}{20 (22.2\%)} & \multicolumn{1}{l|}{31 (36.0\%)} & \multicolumn{1}{l|}{12 (28.6\%)}  & 51 (28.7\%)  \\ \hline
DC8                                   & \multicolumn{1}{l|}{57 (25.9\%)}                     & \multicolumn{1}{l|}{10 (22.7\%)} & \multicolumn{1}{l|}{26 (28.9\%)} & \multicolumn{1}{l|}{21 (24.4\%)} & \multicolumn{1}{l|}{11 (26.2\%)}  & 46 (25.8\%)  \\ \hline
\end{tabular}
\caption{\Modified{Counts and percentages of survey respondents' choices of access methods, sources of enjoyment during visual arts appreciation, motivation, and design considerations across two demographic factors: vision level and onset of blindness. TB = Totally blind, LP = Some light perception, LB = Legally blind, C = Congenitally blind, A = Acquired blindness later. DC1 = Have the option to access the art piece through multiple modalities (such as tactile and audio) at the same time, DC2 = Have visual details delivered through tactile and background information delivered through audio, DC3 = Have a clear art description standard discussed and agreed with describers-such as what vocabularies to use, DC4 = Have flexibility over the pace \& location \& time of art appreciation experience, DC5 = Have a social setting where you feel more comfortable and less stigmatized to enjoy visual art, DC6 = Have the ability to filter out subjective interpretation in the art description-such as those from sighted peers or docents, DC7 = Have art descriptions adapted to your experience and knowledge-such as skipping information that you already knew, DC8 = Have the ability to acquire answers from someone you trust whenever you have a question.}}
\label{table:table_sum}
\end{table}

\newpage

\section{Statistical Test Results}

% \usepackage{multirow}
\begin{table}[h]
\begin{tabular}{|l|ll|ll|}
\hline
\multirow{2}{*}{}                     & \multicolumn{2}{l|}{\textbf{Vision Level}} & \multicolumn{2}{l|}{\textbf{Onset of Blindness}} \\ \cline{2-5} 
                                      & \multicolumn{1}{l|}{$\chi^2$}        & p         & \multicolumn{1}{l|}{$\chi^2$}           & p            \\ \hline
\textbf{Preferred Methods}            & \multicolumn{1}{l|}{}          &           & \multicolumn{1}{l|}{}             &              \\ \hline
Recorded audio at art venues          & \multicolumn{1}{l|}{3.77}      & 0.152     & \multicolumn{1}{l|}{0.65}         & 0.419        \\ \hline
Tactile access at art venues          & \multicolumn{1}{l|}{5.84}      & 0.054     & \multicolumn{1}{l|}{10.56}        & 0.001        \\ \hline
Sighted assistance at art venues      & \multicolumn{1}{l|}{0.07}      & 0.968     & \multicolumn{1}{l|}{2.68}         & 0.101        \\ \hline
Smart device assistance at art venues & \multicolumn{1}{l|}{0.71}      & 0.702     & \multicolumn{1}{l|}{4.26}         & 0.039        \\ \hline
Access visual arts online             & \multicolumn{1}{l|}{8.5}       & 0.014     & \multicolumn{1}{l|}{7.37}         & 0.007        \\ \hline
\textbf{Source of Enjoyment}          & \multicolumn{1}{l|}{}          &           & \multicolumn{1}{l|}{}             &              \\ \hline
Imagination                           & \multicolumn{1}{l|}{1.78}      & 0.41      & \multicolumn{1}{l|}{0.23}         & 0.629        \\ \hline
Touching                              & \multicolumn{1}{l|}{9.17}      & 0.01      & \multicolumn{1}{l|}{4.94}         & 0.026        \\ \hline
Learning related stories              & \multicolumn{1}{l|}{2.3}       & 0.317     & \multicolumn{1}{l|}{0.12}         & 0.729        \\ \hline
Conversation with peers               & \multicolumn{1}{l|}{2.29}      & 0.319     & \multicolumn{1}{l|}{0}            & 1            \\ \hline
Physically co-present with art pieces & \multicolumn{1}{l|}{6.04}      & 0.049     & \multicolumn{1}{l|}{1.4}          & 0.236        \\ \hline
\textbf{Motivation}                   & \multicolumn{1}{l|}{}          &           & \multicolumn{1}{l|}{}             &              \\ \hline
Learning                              & \multicolumn{1}{l|}{2.02}      & 0.364     & \multicolumn{1}{l|}{4.97}         & 0.026        \\ \hline
Encouragement                         & \multicolumn{1}{l|}{6.93}      & 0.031     & \multicolumn{1}{l|}{0.63}         & 0.428        \\ \hline
Activism                              & \multicolumn{1}{l|}{0.41}      & 0.813     & \multicolumn{1}{l|}{0}            & 1            \\ \hline
Social                                & \multicolumn{1}{l|}{2.8}       & 0.246     & \multicolumn{1}{l|}{2}            & 0.158        \\ \hline
Relaxation                            & \multicolumn{1}{l|}{2.19}      & 0.335     & \multicolumn{1}{l|}{0}            & 1            \\ \hline
Nostalgic                             & \multicolumn{1}{l|}{6.98}      & 0.031     & \multicolumn{1}{l|}{3.43}         & 0.064        \\ \hline
\textbf{Desired Design Changes}       & \multicolumn{1}{l|}{}          &           & \multicolumn{1}{l|}{}             &              \\ \hline
DC1                                   & \multicolumn{1}{l|}{0.79}      & 0.674     & \multicolumn{1}{l|}{1.89}         & 0.169        \\ \hline
DC2                                   & \multicolumn{1}{l|}{1.71}      & 0.424     & \multicolumn{1}{l|}{1.19}         & 0.276        \\ \hline
DC3                                   & \multicolumn{1}{l|}{7.23}      & 0.027     & \multicolumn{1}{l|}{1.01}         & 0.314        \\ \hline
DC4                                   & \multicolumn{1}{l|}{1.04}      & 0.595     & \multicolumn{1}{l|}{0.42}         & 0.517        \\ \hline
DC5                                   & \multicolumn{1}{l|}{0.94}      & 0.626     & \multicolumn{1}{l|}{2.77}         & 0.096        \\ \hline
DC6                                   & \multicolumn{1}{l|}{0.76}      & 0.685     & \multicolumn{1}{l|}{1.2}          & 0.274        \\ \hline
DC7                                   & \multicolumn{1}{l|}{4.16}      & 0.125     & \multicolumn{1}{l|}{0}            & 1            \\ \hline
DC8                                   & \multicolumn{1}{l|}{0.75}      & 0.688     & \multicolumn{1}{l|}{0}            & 1            \\ \hline
\end{tabular}
\caption{\Modified{Summary of statistical results from N=220 survey respondents' preferred access methods, source of enjoyment, motivation, and design considerations across onset of blindness and vision level. DC1 = Have the option to access the art piece through multiple modalities (such as tactile and audio) at the same time, DC2 = Have visual details delivered through tactile and background information delivered through audio, DC3 = Have a clear art description standard discussed and agreed with describers-such as what vocabularies to use, DC4 = Have flexibility over the pace \& location \& time of art appreciation experience, DC5 = Have a social setting where you feel more comfortable and less stigmatized to enjoy visual art, DC6 = Have the ability to filter out subjective interpretation in the art description-such as those from sighted peers or docents, DC7 = Have art descriptions adapted to your experience and knowledge-such as skipping information that you already knew, DC8 = Have the ability to acquire answers from someone you trust whenever you have a question.}}
\label{test_summary}
\end{table}

\newpage

\section{Onset of Blindness on Preferred Method}

\begin{figure}[h]
    \centering
    \includegraphics[width=1\columnwidth]{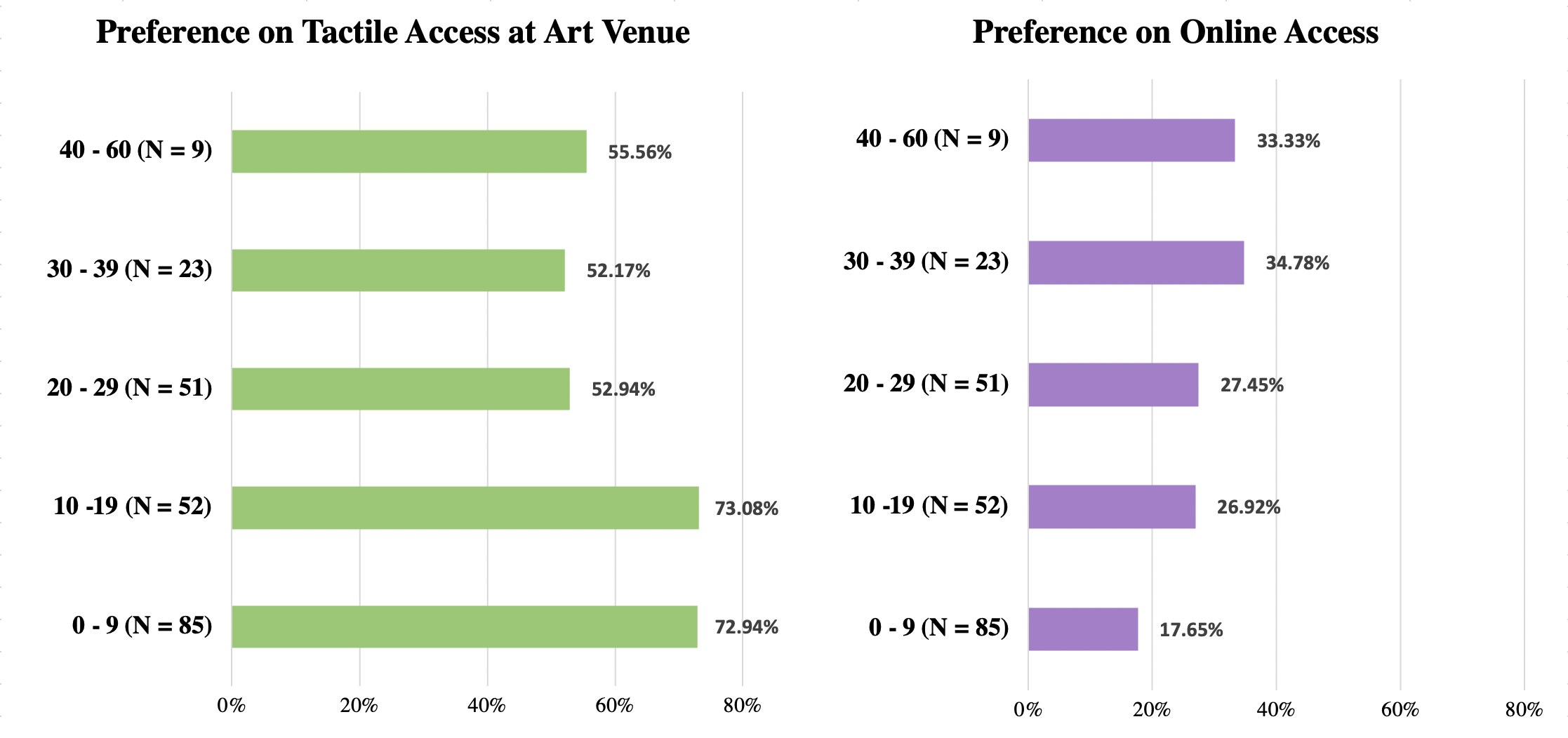}
    \caption{\Modified{Survey respondents’ preferences on tactile access at art venue (left) and online access (right) across different age range of blindness onset.}}
    \label{fig:onset_age}
    \Description{Two bar charts, the left chart shows the preference on tactile at art venue. The y-axis has five different age groups (40-60, 30-39, 20-29, 10-19, 0-9) and the x-axis has different percentage rates. The right figure shows the preference on online access. The y-axis has five different age groups (40-60, 30-39, 20-29, 10-19, 0-9) and the x-axis has different percentage rates.}
\end{figure}

% \section{Survey Results on Motivations and Vision Conditions}

% \begin{figure}[h]
%     \centering
%     \includegraphics[width=1\columnwidth]{Figures/fig_motivation.png}
%     \caption{\Modified{Survey respondents' choices of motivation for visual arts appreciation across onset of blindness (right) and vision level (left).}}
%     \label{fig:motivation}
%     \Description{Two bar charts, the left chart shows the motivations vs. onset of blindness (congenital, acquired). The y-axis has six different categories (learning, relaxation, social, encouragement, nostalgic, and activism) and the x-axis has different percentage rates. The right chart shows the motivations vs. vision level (totally blind, some light perception, totally blind). The y-axis has six different categories (learning, relaxation, social, encouragement, nostalgic, and activism) and the x-axis has different percentage rates.}
% \end{figure}

\end{document}